\documentclass[11pt,a4paper]{article}
\usepackage{jcappub}
\usepackage{graphicx}%  Default Latex 2eps package for embedding figures
\usepackage{rotating}
\usepackage{epsfig}
\usepackage[lofdepth,lotdepth,caption=false]{subfig}
\usepackage{bm}
\usepackage{amsmath}
\usepackage{graphicx}% Include figure files
\usepackage{dcolumn}% Align table columns on decimal point
\usepackage{bm}% bold math

\usepackage{url}
\usepackage{amsmath,amssymb,graphicx}
\usepackage{amsmath}
\usepackage{amssymb}
\usepackage{mathrsfs}
\usepackage{accents}
\usepackage{epsfig}

\usepackage{mathtools}
\usepackage{mathrsfs}

\usepackage{color}
\begin{document}

\title{Impact of standard neutrino oscillations and systematics in proton lifetime measurements}

\author[a,b,1]{D.R. Gratieri,\note{Corresponding author.}}
\author[b]{M.M.  Guzzo}
\author[b]{and O.L.G. Peres}

\affiliation[a]{EEIMVR, Universidade Federal Fluminense, RJ, Brazil}
\affiliation[b]{Instituto de F\'isica Gleb Wataghin - UNICAMP, {13083-859}, Campinas SP, Brazil}

\date{\today}

%\begin{abstract}

\abstract{We use atmospheric neutrino phenomenology to obtain the expected background to proton decay in large underground neutrino detectors, like DUNE. We introduced, for the first time in this kind of analysis, the experimentally confirmed neutrino oscillations of the atmospheric neutrino observations which reduce by a factor of 40\% the corresponding background for nucleon decay channel $p\rightarrow \mu^{+}+\pi^{0}$. Furthermore, we infer the impact of four systematics on such background: the overall efficiency, the muon reconstruction energy resolution, the resonant neutral pion cross-section and the neutral pion angular resolution. Considering a 40 kton detector with efficiency 45\%, our analysis leads to an error band in the lower limit for the proton lifetime, from $7.9 \times 10^{33}$~years to {\bf $1.1 \times 10^{34}$}~years at $90\%$~C.L.. These numbers can be compared with the current mode dependent experimental limits $\tau >  10^{31}-10^{33}$~years at $90\%$~C.L.. Finally we investigate how this limit can be further improved with the enhancement of the efficiency of the experiment which can be obtained, for instance, with the implementation of the ARAPUCA device in DUNE.}

%\pacs{14.60.Lm, 14.60.St, 14.60.Pq}

\maketitle

\section{Introduction}
The conservation of barion number is an accidental symmetry of the Standard Model~\cite{ref:GSW,ref:QCD} in the sense that among all the possible theoretical scenarios, nature itself seems to preserve 
the number of bound states formed by three valence quarks or anti-quarks in all processes involving elementary particles. An opposite picture happens in meson case. Bound states of a valence quark and anti-quark can be produced without 
any penalties of all known conservation rules. 
Even though, several well established theories predict baryon violation and consequently nucleon radiative decay into lighter subatomic particles, mainly into a lepton (electrons or muons) and a meson (pions or kaons)\footnote{See \cite{ref:pavel} for a review on this subject.}. Different theories predict different nucleon lifetimes, some of them already discarded by the present mode dependent experimental limit $\tau > 10^{31}-10^{33}$~years at $90\%$~C.L.~\cite{ref:PDG}. Minimal $SU(5)$ Grand Unification~\cite{ref:GSW-1}, for instance,  foresees proton lifetime in the range already experimentally discarded $\tau= 10^{30}-10^{31}$~years. A second group of theories predicts proton lifetime just in the range of the experimental measurements, like as Minimal SUSY SU(5) ($\tau = 10^{28}-10^{34}$~years~\cite{ref:11,ref:12,ref:13}), SUGRA SU(5) ($\tau = 10^{32}-10^{34}$~years~\cite{ref:14,ref:15}), SUSY SO(10) ($\tau = 10^{32}-10^{35}$~years~\cite{ref:16,ref:17,ref:18,ref:19,ref:21}), Minimal non-SUSY SU(5) ($\tau = 10^{31}-10^{38}$~years~\cite{ref:24}). While a third group of theories foresees the proton lifetime $\tau > 10^{34}$~years, well above the experimental limits, like as SUSY SU(5) or SO(10) in 6 dimensions~\cite{ref:18}, Flipped SU(5)~\cite{ref:22}, Split SU(5) SUSY~\cite{ref:23}, SU(5) in 5 dimensions~\cite{ref:25,ref:26}, GUT-like models from Type IIA Strings~\cite{ref:27}. 
The aim of this article is to investigate the potential of large deep underground neutrino detectors like the {\it Deep Underground Neutrino Detector} (DUNE)~\cite{ref:dune-whitep}, to improve the current experimental proton lifetime limit. This can be particularly interesting for testing and, eventually, discard, the theories of the second group listed above. 

The quest of finding proton decay is fully related with atmospheric neutrinos since they consist the main background to the most probable proton decay channels. The need of determination of such background leaded to the {\it Atmospheric Neutrino Problem}, a deficit of order of $50\%$ found in  number of {\it muon-like} events produced in up-going  direction  in 
Super-Kamiokande detector~\cite{ref:SK-evidence}. The solution of this problem, combined with the solution of  {\it Solar Neutrino Problem}~\cite{ref:Sudbury}, favored the scenario of  neutrino flavor oscillation picture induced by mass-mixing neutrino formalism~\cite{ref:Pontecorvo,ref:MNS}. Neutrino oscillations are also necessary to describe solar neutrinos~\cite{ref:gallex,ref:sage,ref:SK-SOLAR,ref:cleveland}, reactor experiments~\cite{ref:kamland}, accelerator experiments~\cite{ref:minos,ref:dayabay} and atmospheric neutrinos~\cite{ref:soudan,ref:macro,ref:atm-ICecube} as well. It is a notable fact that Standard Neutrino Oscillations can describe successfully all these experiments where neutrinos and anti-neutrinos from different sources, different flavors  and energies, travel very distinct distance and cross different matter potentials. No other  theoretical mechanism  survived to these phenomenological constraints. As pointed in Ref.~\cite{ref:DUNE-OS}, the standard oscillation mechanism is now well established and all parameters can be measured with high level of accuracy. In this article, we include such accuracy in the determination of atmospheric background  for proton decay.

The huge lower limits on proton lifetime makes the massive underground detectors, like Super-Kamiokande (SK)~\cite{SK-pdecay} and its full extension to the Hyper-Kamiokande (Hyper-K)~\cite{ref:hyperK}, perfect places to look for proton decay. Furthermore, DUNE~\cite{ref:dune-whitep}, an international facility in which the world most intense neutrino beam will be produced in Fermilab accelerator and will strike  a $40$~kton made of liquid argon detector in the Stanford Underground Research Facility, which is $\approx 1300$~km far from the neutrino source, includes in 
its scientific program  proton decay measurements. Indeed, it consists in the most promissory experiment to probe some of the proton decay channels~\cite{ref:dune-whitep}. Hence, the most exciting scenario to proton decay searches would be possible in near future, with both Hyper-K and DUNE experiments running. 

DUNE  will use the {\it Liquid Argon} drift technology~\cite{ref:lqdar40} to detect neutrinos. The use of this new process of neutrino detection demands  the specific determination of the atmospheric neutrino background to proton decay. In Ref.~\cite{ref:rubia} an extensive calculation of this background was performed, including all theoretical channels to proton decay available at that time. It includes atmospheric neutrino background as well as the one due to atmospheric muons. In comparison to Ref.~\cite{ref:rubia}, in this work we focus our attention in the specific decay  channel:
\begin{equation}
 p\rightarrow \mu^{+}+\pi^{0}\rightarrow \mu^{+}+\gamma+\gamma~,
 \label{eq:pdec01}
 \end{equation}
 and add physical effects we think cannot be disregarded in the determination of atmospheric neutrino background to such process. They are:
\begin{itemize}
\item The Standard Neutrino Oscillations and
\item Resonant neutral pion production cross-section is tunned to MiniBooNE~\cite{AguilarArevalo:2010zc} data.
\end{itemize}

Neutrino cross-sections play the most important rule in the determination of atmospheric neutrino events at any experiment. Atmospheric neutrino fluxes are peaked at neutrino energies around $1.0$~GeV~\cite{ref:Honda:2006qj} and in this region the three main neutrino-nucleus scattering process {\it (Quasi-Elastic(QE), Deep Inelastic Scattering(DIS), and Resonant Pion Production}) are present~\cite{ref:zeller}. Also, the kinematic range for the products in 
Eq.~(\ref{eq:pdec01}) is largely constrained. This means that the phase space for a muon event due to atmospheric neutrinos  lies in the  energy window that count as background to Eq.~(\ref{eq:pdec01}) is very small.  As will be discussed in Section \ref{sec:bkgd}, this implies in knowing the resonant pion production cross-section in a regime of very low transfered momentum,  $Q^{2}< 1.0$~GeV$^{2}$,  and large values of Bjorken variable, $x_{Bj} \rightarrow 1$.  At such energies,  the formalism to resonant pion production  must take into account nuclear effects, both in first vertex interaction as well in the Final State Interactions (FSI)\cite{ref:FSI} of leptons and  pions with the nuclear medium they are produced.  

The paper is organized as follows. In 
Section~\ref{sec:bkgd} we define the  reaction  responsible for the background and develop the kinematics of such process. In Section~\ref{sec:cs} we introduce the resonant charged current 
neutrino-nucleon pion production cross-section and determine its behavior in the kinematic region of interest here. Section~\ref{sec:SO} is devoted to the introduction of  standard neutrino oscillations. In Section~\ref{sec:Natm} we define the calculation of background.  Results are in Section~\ref{sec:results} and our conclusions are presented in Sec.~\ref{sec:conc}.

\section{Kinematics of proton decay signal and background }
\label{sec:bkgd}
In this work  we are interested in the background due to  atmospheric neutrinos to the proton decay signal for the decay reaction Eq~(\ref{eq:pdec01}).  Assuming that protons are at rest when decaying,  $\vec p_{\mu}$ is given by Eq.~(\ref{pmu}). Taking into account that protons are not free inside the detector, but bounded to other protons and neutrons in the nucleons, the effective proton mass must be calculated~\cite{ref:testeSK}. When we assume $m_{p}=925$~MeV/c$^{2}$ in Eq.~(\ref{pmu}), we find for the muon momentum $|\vec p_{\mu}|=446.4$~MeV/c.  In  general case, protons should have some initial momentum before decaying.  After some algebra (see Appendix \ref{apk})  the desired relation for muon momenta in this case is given in Eq~(\ref{pmuslab1})~\cite{goldanski}. Clearly, the maximum (minimum) of $\vec{p}_{\mu}$ occurs when the muon emission is in the same(opposite) direction of proton's initial momentum,  $\theta_{\mu}=0$ ($\theta_{\mu}=180^{0}$), where $\theta_{\mu}$ is the emission angle of the muon with respect to the proton momentum in the laboratory frame.  In this case, from Eq.~(\ref{pmuslab0}),  $ p_{\mu,\rm max(min)}=588.7(340.7)~{\rm MeV}$. Here we assume the maximum momentum for the proton to be $p_{p}=250$~MeV. From~\cite{ref:icarus}, in liquid argon technology,   the muon momenta resolution is  of order of $\delta p_{\mu}/p_{\mu}\approx 18\%$ which implies in a  $  61.3 (105.6)$~MeV resolution band which determines the energy and momenta for the products of Eq.(\ref{eq:pdec01}). Explicitly we have, 
\begin{equation}
279\le p_{\mu}\le 694~{\rm MeV}~,~~~~~~~~~~~~~~~~~~~~292\le E_{\mu}\le 702~{\rm MeV}~.
\label{pwindow}
\end{equation}

At same time, Eq (\ref{pwindow}) informs us the width of energy and momentum window in which the muons and pions produced by atmospheric neutrino interaction will count as background to proton decay. In Sec. \ref{sec:Natm} we detail the procedure to calculate the background to Eq. (\ref{eq:pdec01}).  Here  we simply introduce  the main (anti)neutrino reaction that generates the background, 
\begin{equation}
\nu_{l}+N\rightarrow l^{-}+\Delta^{0}\rightarrow l^{-}+P+m\pi^{0};~~~~\bar\nu_{l}+P\rightarrow l^{+}+\Delta^{0}\rightarrow l^{+}+N+m\pi^{0}~.
\label{pires1}
\end{equation}

\begin{figure}[hbt]
\begin{center}
 \includegraphics[scale=0.4]{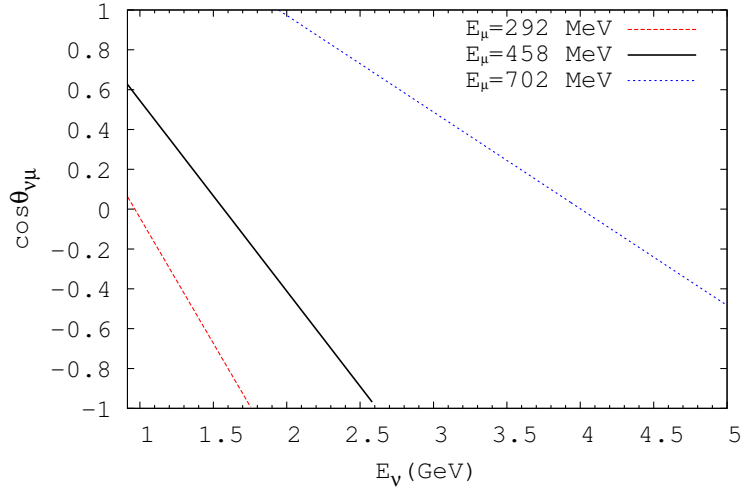}
\caption{The possible values for $\cos(\theta_{\nu\mu})$ for different values of $E_{\mu}$ as function of neutrino energy.}
 \label{cthmu}
 \end{center}
 \end{figure}

In both cases, signal and background, to constrain neutral pion one can use the fact that in liquid argon technology the energy resolution for EM-showers is of $3\%\sqrt{E({\rm GeV})} $~\cite{ref:lariat00}. For $\pi^{0}\rightarrow \gamma \gamma\rightarrow e^{+}e^{-}+e^{+}e^{-}$ the approximate resolution is $\delta E_{\pi^{0},{\rm min} }\approx 34$~MeV and $\delta E_{\pi^{0},{\rm max}}\approx 48$~MeV. For background calculation, this  implies that $\approx 38\%$ of produced neutral pion in Eq.~(\ref{pires1}) will follow in the range of momenta that counts as background~\cite{ref:nutini}. Also,  the muon final state energy to be inside the limits given in Eq.~(\ref{pwindow}) implies in constrains to the kinematic variables. For the angle between incoming neutrino and
outgoing muon  $\theta_{\nu\mu}$ we have
\begin{equation}
 \cos(\theta_{\nu\mu})=\dfrac{m^{2}_{\Delta}-m^{2}_{\rm n}-m^{2}_{\mu}+2m_{n}E_{\mu}+2(E_{\mu}-m_{\rm n})E_{\nu}}{2p_{\mu}E_{\nu}}.
\end{equation}
In Fig.~\ref{cthmu} we show the allowed values for the scattering angle as function of $E_{\nu}$ for the cases of proton at rest, as well as for the limits in $E_{\mu}$ given form Eq.~(\ref{pwindow}). 

The above  kinematic cuts from Eq.~(\ref{pwindow}) also constrain the transfered momentum from the incoming neutrino to the hadronic system  in Eq.~(\ref{pires1}), given from~\cite{ref:Predazzi}, 
\begin{equation}
Q^{2}= m^{2}_{\rm n}-m^{2}_{\Delta}+2m_{n}(E_{\nu}-E_{\mu}). 
\label{Q2}
\end{equation}
Such constrains also are applied  to the  Bjorken variable, which is defined as the fraction of proton momentum carried by each parton, 
\begin{equation}
x_{\rm Bj}=\dfrac{Q^{2}}{2m_{\rm n}(E_{\nu}-E_{\mu})}.  
\end{equation}
\begin{figure}[hbt]
\begin{center}
%\hspace{-1.cm}
\includegraphics[scale=0.4]{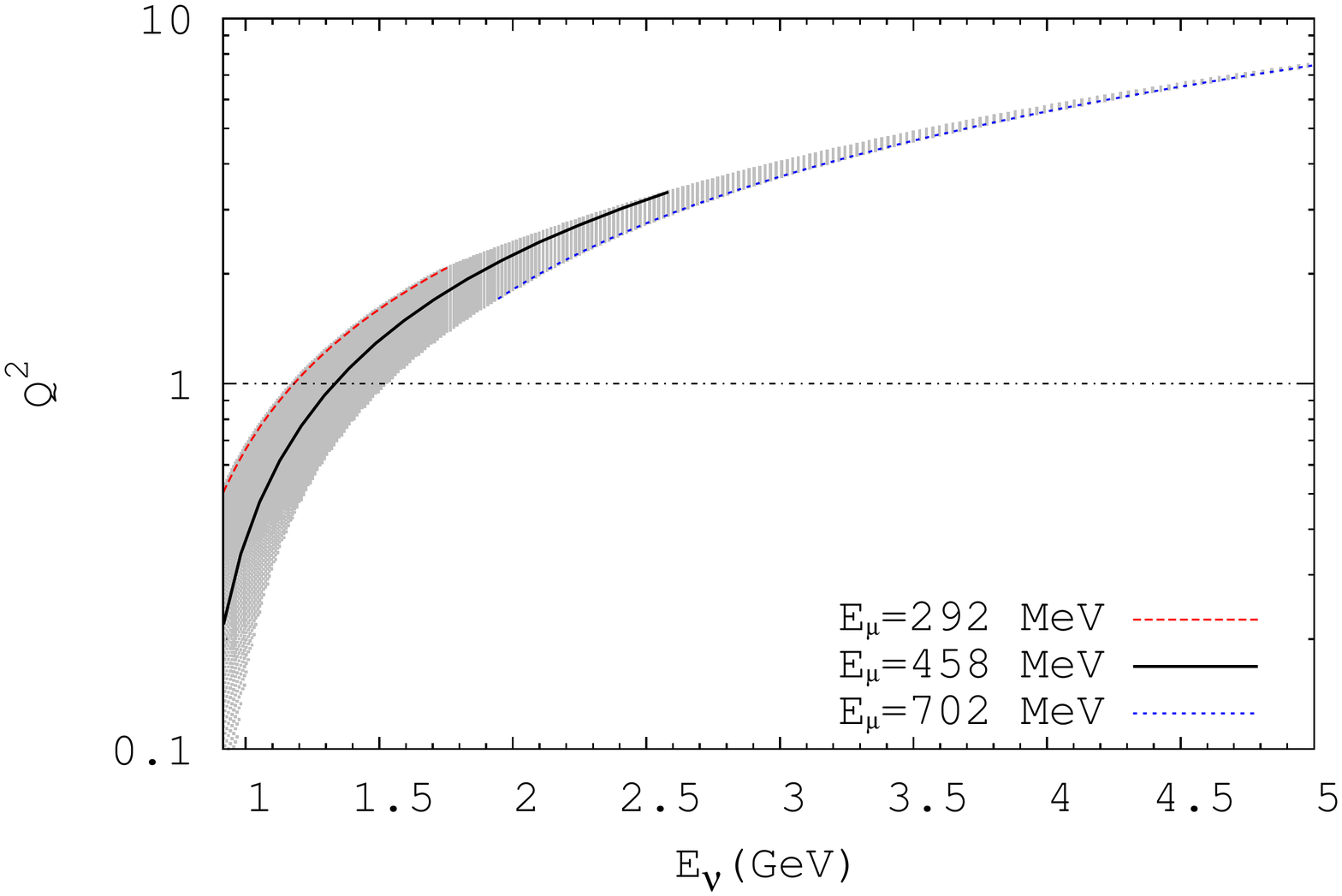}\vspace{-1.75cm} 

\includegraphics[scale=0.4]{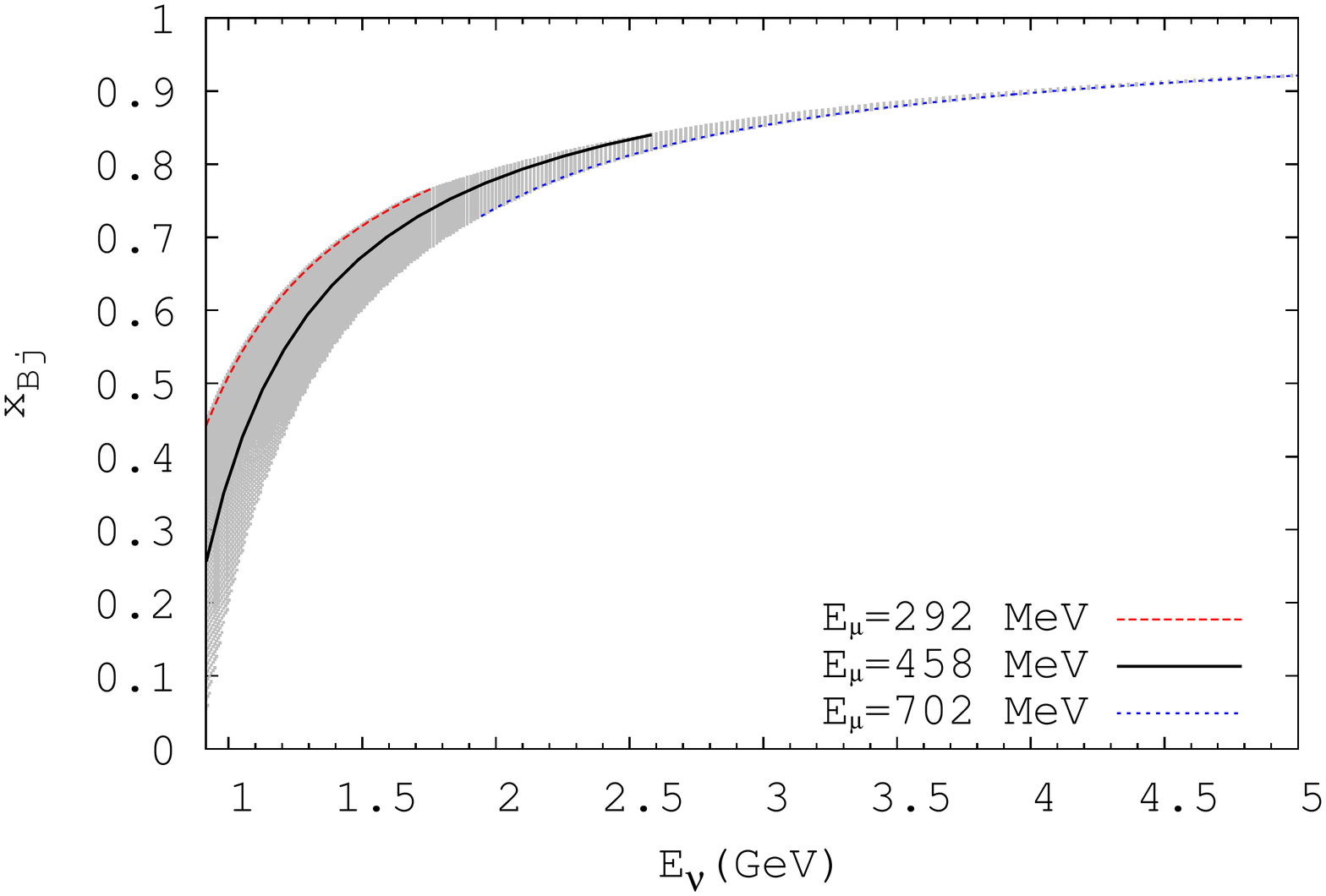}
\caption{Upper panel: Allowed values for $Q^{2}$ in 
Eq~.(\ref{pires1}) for different values of $E_{\mu}$ as function of neutrino energy. 
Lower panel: The same  for $x_{\rm Bj}$.}
 \label{Q2-fig}
 \end{center}
 \end{figure}
 
In the upper panel of Fig.~\ref{Q2-fig} we show the possible values from $Q^{2}$ in Eq.~(\ref{pires1}) as a function of neutrino energy. The values given from the minimum  and  maximum values of $E_{\mu}$ as well as for the case of proton at rest are indicated.  The main point here is that most part of the phase space for the Eq.~(\ref{pires1}) lies in the $Q^{2}\le 1.0$~GeV$^{2}$.  In Fig.~\ref{Q2-fig}, lower panel, shows our results for the allowed space for $x_{Bj}$ as function of neutrino Energy. The minimum  and  maximum values of $E_{\mu}$ as well as for the case of proton at rest are indicated.  Hence, the region in which the reaction Eq. (\ref{pires1}) is allowed reduces the phase space for the reaction but also implies  in the large $x_{Bj}$ and  includes $Q^{2}\le 1.0$~GeV$^{2}$ region. This determines the dynamic description of nucleon structure.  Such region  is not included in most part of parton density functions found in literature~\cite{ref:PDG}. See more in  Sec~\ref{sec:cs}.

\section{Cross-sections for the atmospheric neutrino resonant interaction}
\label{sec:cs}

 Here we address the cross-sections that are important in atmospheric neutrino interactions. The main source of systematic error in the calculation of number of events due to atmospheric neutrino is the neutrino-nucleon  cross-section. Atmospheric neutrino flux~\cite{ref:Honda:2006qj} extends from few MeVs to several hundred GeVs.   In such  kinematic region there are three main Charged Current (CC) channels for the process do occur,  quasi-elastic (QE),  resonant scattering (RES) and  Deep Inelastic Scattering (DIS). All these processes are present and there is no separation line between them. Also the description of target nucleus modifies as the energy of projectile increases. 
 
 At low neutrino energies, around few hundred  MeVs, neutrino interaction is given through QE scattering, in which the target nucleus remains approximately unaffected. The low transfered momentum implies in large wavelength associated with bosons $W^{\pm}$,  and hence the neutrino-nucleon interaction is described in terms of form factors~\cite{ref:Smith, Strumia:2003zx}. On the higher energy limit, $E_{\nu}>$ few GeVs, the energy transfered to the target nucleus is so high that nucleus fragmentation occurs and a continuous spectrum of hadronic states is created. This is the signature of DIS process. At this kinematic regime, due to high transfered momentum from projectile to target, the virtual $W^{\pm}$ as small  wavelength and couples to quarks. In this case Parton Density Functions are needed to take into account parton content inside the nucleus. At such high energy limit, perturbative Quantum ChromoDynamics (pQCD) can be used to describe the interaction. See~\cite{ref:Predazzi} for a review. 
 
\begin{figure}[hbt]
\begin{center}
\includegraphics[scale=0.4]{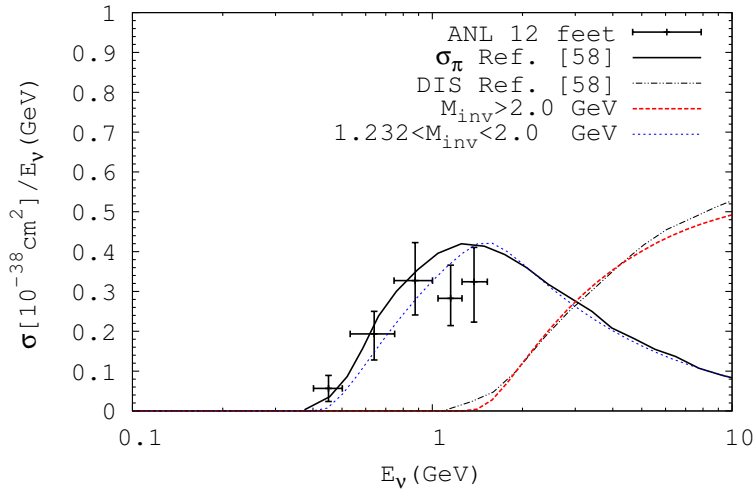}
\caption{ Our results for the DIS and resonant neutrino scattering as given from Eq. (\ref{defdis}) for different cuts in $M_{\rm inv}$ are compared with $\sigma_{\rm \pi}$ and $\sigma_{\rm DIS}$  from~\cite{Morfin:2012kn} and with pion production data from~\cite{ref:aln12}. }
\label{fig:numuQEII}
\end{center}
\end{figure}

 At middle  neutrino energies however, the typical transfered momentum to a quark in the interaction is not enough to break the target, but makes quarks occupy a higher energy orbital. This new bound state of three quarks is known as a nucleon resonance, and between all the possible spectra, the lighter one is the $\Delta(1232)$. Such resonance decays in a typical strong force process, $\Delta \rightarrow N+m\pi $, with $\tau \approx 10^{-23}$~s.  The hole process is then called neutrino resonant pion production (single pion production if $m=1$ ) and  is represented in  Eq.~(\ref{pires1}) for the case of $\Delta^{0}$ production. Indeed, in the energy region of $1.0\le  E_{\nu} \le 5.0$~GeV, there is the necessity  to interpolate  resonance and pQCD results~\cite{Mosel:2016cwa}. Moreover, the quark-hadron duality idea~\cite{ref:Poggio,ref:Shifman} states that hadronic cross-sections  like DIS, when averaged in energy, do agree with cross-sections from pQCD (in terms of quarks and gluons). Hence, one could calculate the cross-section we need as neutrino scattering a nucleon, which requires the knowledge of hadronic resonance production form factors, or as neutrino scattering to a quark inside the nucleon, and hence Parton Density Functions are needed in all phase space allowed to the reaction. Following the later approach, the formalism to describe DIS process we apply is defined in~\cite{ref:Predazzi}. We include  in  DIS process the dependence of charged lepton masses $m_{l}$, $\{l=e,~\mu,~\tau\}$ leading to
\begin{eqnarray}
\frac{d\sigma}{dxdy}&=& \frac{G^{2}_{F}M}{\pi}E_{\nu}\left(\frac{M^{2}_{W}}{Q^{2}+M^{2}_{W}} \right)^{2} \left\{y\left(xy- \frac{m^{2}_{l}}{2E_{\nu}M} \right)F_{1}\right.\nonumber\\
&+&\left. \left( 1 -y  \frac{Mxy}{2E_{\nu}} -\frac{m^{2}_{l}}{4E^{2}_{\nu}}\right)F_{2}\right.\nonumber \\  
&\pm&\left.  \left( xy \left(1-\frac{x}{y}\right) -y\frac{m^{2}_{l}}{4ME_{\nu}} \right)F_{3}  +\left(xy \frac{m^{2}_{l}}{2ME_{\nu}} +\frac{m^{4}_{l}}{4M^{2}E^{2}_{\nu}}\right)F_{4} \right.\nonumber\\
&-&\left. \frac{m^{2}_{l}}{2ME_{\nu}}F_{5}       \right\},         
\label{defdis}
\end{eqnarray}
where $y=\dfrac{E_{\nu}-E_{l}}{E_{\nu}}$, and  $F_{i}$ are the partons density  functions. The $(-)+$ signal refers to (anti)neutrino scattering. In leading order approach they are
\begin{eqnarray}
2xF_{1}&=&F_{2}, \nonumber \\
F_{4}&=&0,  \nonumber \\
xF_{5}&=&F_{2}~,\nonumber \\
\label{DIS1}
\end{eqnarray}
that depend on  the variables $x, Q^{2}$ and obtained from accelerator data fitting. Several  groups in literature  specify  their PDF's as result of fitting colider experiments, and also, the region in $x,Q^{2}$ of validity of it.  Moreover, in literature, the  DIS region was defined by $M_{\rm inv}>2.0$~GeV, here $M_{{\rm inv}}$ is the mass of final hadronic state. However, we stress here the there is no clear line to separate DIS from the resonant pion production. It depends on the way you count the hadronic resonances. The theoretical formalism to  resonant pion production is due to~\cite{Fogli:1979cz}. The DIS formalism given in Eq.(\ref{defdis}) recovers to pion production when we apply the appropriate cuts in mass of final hadronic state, $1.232\le M_{\rm inv}\le 2.0$~GeV. In Fig.~\ref{fig:numuQEII}, we show the predictions from Eq.~(\ref{defdis}) for DIS and pion resonant production,  in the  isoscalar case, and compare it with the  reference~\cite{Morfin:2012kn}. Also are shown the experimental values indicated from~\cite{ref:aln12}.  We obtain  good  agreement with the same reference
for both process just setting the proper cuts in $M_{\rm inv}$ discussed above. We use the old GRV-94~\cite{Gluck:1994uf} since it has a lower threshold for the transfered momentum, i.e. $Q^{2} \le 0.3$~GeV$^{2}$.  We also compare  it with predictions from CTEQ5~\cite{Lai:1999wy} and  MST~\cite{ref:MST} parton distribution functions and find no greater difference for 
$E_{\nu} \leq 7 $~GeV. Our choice is based in the low $Q^{2}$ limit of validity of such distributions. Hence, apart from the uncertainty from choice of PDF,  such simplistic formalism is enough to describe old bubble chamber data~\cite{ref:aln12}. 
 However, the resonant pion production reveals to be harder to describe.  At such intermediate energies the nuclear effects play an important role and corrections are necessary both from theoretical point of view as well as to describe the new data ~\cite{Abe:2015oar,Abe:2016aoo,ref:k2k,ref:Sciboone,AlcarazAunion:2009ku,ref:minerva,Fiorentini:2013ezn}. These  effects are present  for the primary vertex interaction as well as  for FSI and surely impact proton decay analyses~\cite{ref:dorota}. 
 
 To describe  the first vertex the nuclear PDF(nPDF)~\cite{ref:PDFlib} are then necessary at values of $Q^{2}$ and $x_{Bj}$ shown  in Fig.~\ref{Q2-fig}. This low $Q^{2}$ region is below the cuts commonly applied determination of PDF from accelerator data, since they  are based on DIS processes.  However, in~\cite{ref:LOWq2} an extension to  
 $Q^{2}\geq 0.07$~GeV$^{2}$ based of flavor asymmetric sea of quarks is made. This include non-linear corrections in DGLAP~\cite{ref:DGLAP,ref:DGLAP2} evolution equations for PDFs.  We verified that when using such PDF there is a reduction of at maximum $12\%$ in resonant pion production cross-section when compared with our result in Fig. \ref{fig:numuQEII}. 
 
\begin{figure}[hbt] 
\begin{center}
\includegraphics[scale=0.4]{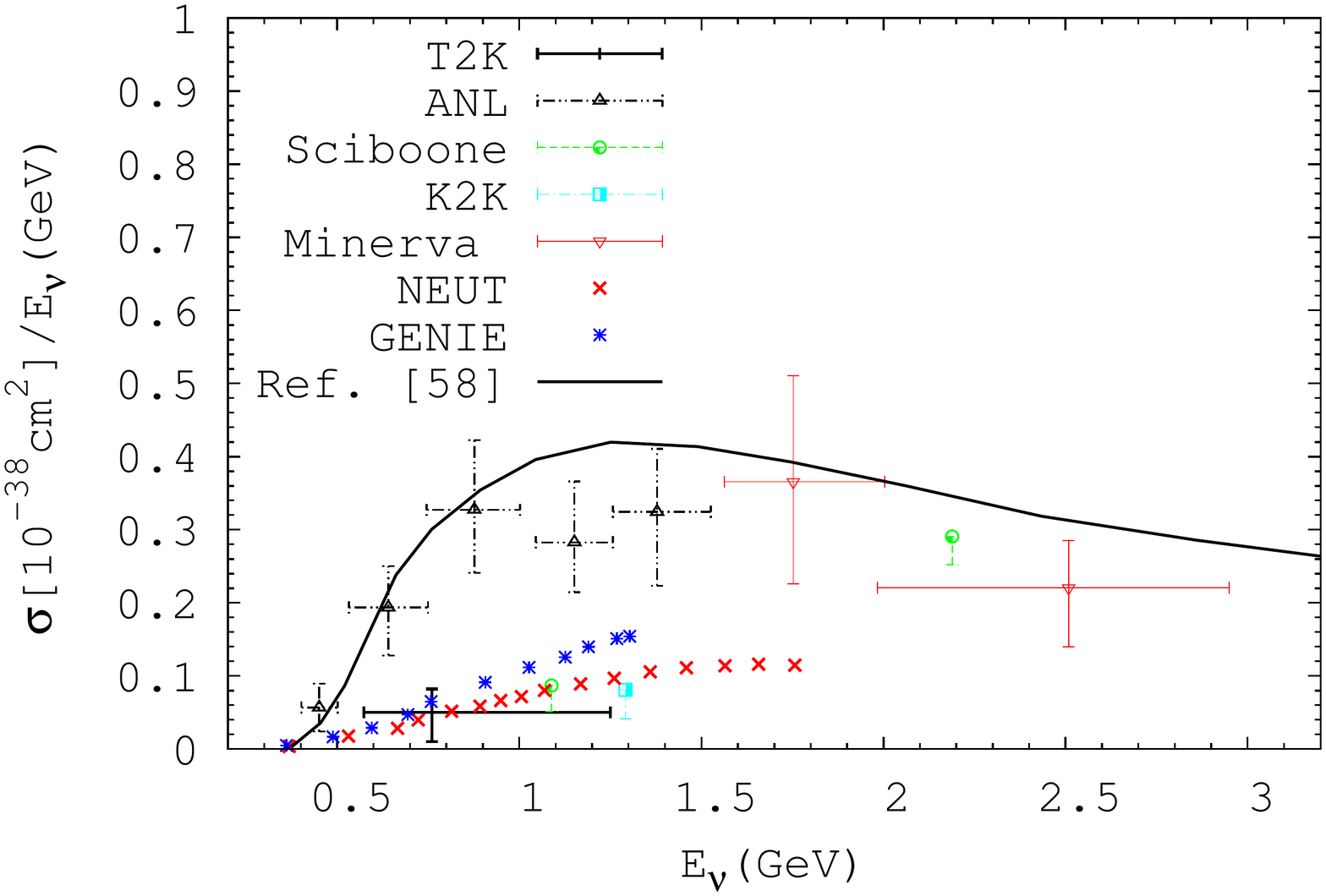}\vspace{-1.5cm}
\includegraphics[scale=0.4]{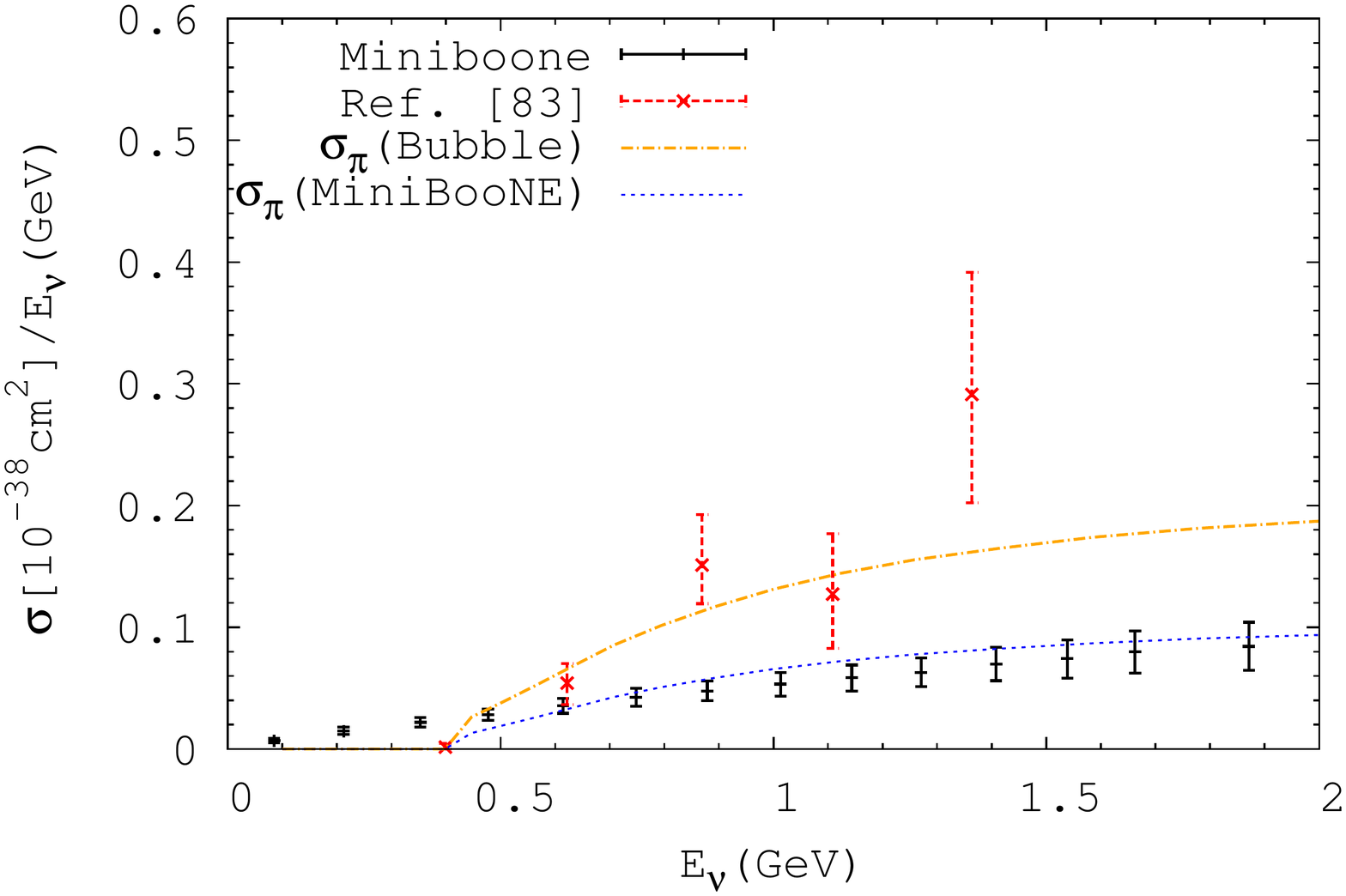}
\caption{Upper Panel: Compilation of resonant charged pion production data and generators predictions from reaction $\nu_{\mu}+N\rightarrow \mu+N'+\pi^{\pm}$.  T2K data and the predictions from GENIE and NEUT event generators are from~\cite{Abe:2016aoo}. Lower Panel: We apply the formalism from ~\cite{Fogli:1979cz} to the process $\nu_{\mu}+N\rightarrow \mu+N'+\pi^{0} $ and describe MiniBooNe and old bubble chamber  data.  Points refers to the data and dashed lines are our results when applied to describe the data set indicated. }
\label{fig:oldnew}
\end{center}
\end{figure}

 The FSI must also be included in resonant pion production cross-section. Two different approaches are found in literature: {\it i-} Uses the effective approach and count the multiple interactions  of produced leptons and pions with nuclear media through the optical theorem, or {\it ii- } Perform a complete microscopic simulation through propagation equations of final state particles inside nuclear medium~\cite{ref:battistoni}. Such procedures are applied in neutrino event generators as NEUT~\cite{NEUT} and GENIE~\cite{Andreopoulos:2009rq}. Both includes the {\it Partially conserved axial current }PCAC models~\cite{Gell-Levy,Rein:2006di,Berger:2008xs}.  In~\cite{Abe:2016aoo,ref:paul} is pointed how generators as GENIE and NEUT show different predictions when using the same model(Rein-Sehgal~\cite{Rein:2006di}, Berger-Sehgal~\cite{Berger:2008xs}, 
 Alvares-Russo~\cite{ref:vacas}), and reciprocally,  different models in same generator also leads to different predictions. Also, in~\cite{Abe:2016aoo} the prediction from GENIE averaged in neutrino  energy is appreciable higher than the data. We show  in Fig.~\ref{fig:oldnew}, upper panel, a comparison of  old and new data for charged pion production, as well   predictions from NEUT and GENIE neutrino events generators as it is shown in~\cite{Abe:2016aoo}. Clearly, there is a discrepancy between the T2K~\cite{Abe:2016aoo},  K2K~\cite{ref:k2k} and  SciBooNe~\cite{ref:Sciboone,AlcarazAunion:2009ku}(low energy point) new data results, and  GENIE and NEUT modern predictions when compared with the old ANL data and models as~\cite{Morfin:2012kn}. However the lower energy points of Minerva~\cite{ref:minerva,Fiorentini:2013ezn} and the higher energy point of  SciBooNe~\cite{ref:Sciboone,AlcarazAunion:2009ku} seems to follow the old pattern.  From new data set is clear the reduction on resonant cross-section for $E_{\nu}<1.5$~GeV. We verified that the inclusion modern PDF which includes low $Q^{2}$ corrections from~\cite{ref:LOWq2}  effectively reduces the resonant cross-section given from Eq. (\ref{defdis}), but not enough to make it compatible with new data. 

For the neutral pion production, we apply the formalism from \cite{Fogli:1979cz} to describe the data from  old Bubble experiments~\cite{ref:barish,ref:maex,ref:maex2,ref:manex3}. We  call $\sigma_{\pi}(Bubble)$  to the resulting cross-section from such process. We also apply such formalism to describe MiniBooNe data~\cite{AguilarArevalo:2010zc}. We call $\sigma_{\pi}(MiniBooNE)$ to the cross-section we obtain in this case. The comparison between such data and the result cross-sections is shown in Fig.~\ref{fig:oldnew}, lower panel.  We clearly see that, except from the very low energy region of MiniBooNE data, which is below the threshold for Eq. (\ref{pires1}), our procedure describe both old and new data. As in the case of charged pion production, the new data to neutral pion production seems to follow a lower pattern when compared with  Bubble experiments. 

As mentioned in Ref.~\cite{Abe:2016aoo}, the reduction in resonant cross-section must have impact on neutrino phenomenology, such as {\it long baseline} and {\it atmospheric neutrinos} experiments. In Super-Kamiokande, since {\it Sub-GeV} data are dominated by low energy neutrinos whose interacts quasi-elastically, such effects on pion production cross-section will be much more relevant  at {\it Multi-GeV} data. By same logic we argue that the  observable most susceptible to it would be precisely  the background to proton decay due to atmospheric neutrinos, for three reasons:

\begin{itemize}
\item It depends almost exclusively on resonant pion production cross section,
\item It is constrained to low neutrino energy range, $E_{\nu}\leq 3.0$~GeV and
\item The exponential decrease with energy of atmospheric neutrino flux makes the region of $E_{\nu}<1.0$~GeV even more important. 
\end{itemize}

\begin{figure}[hbt]
\begin{center}
\includegraphics[scale=0.4]{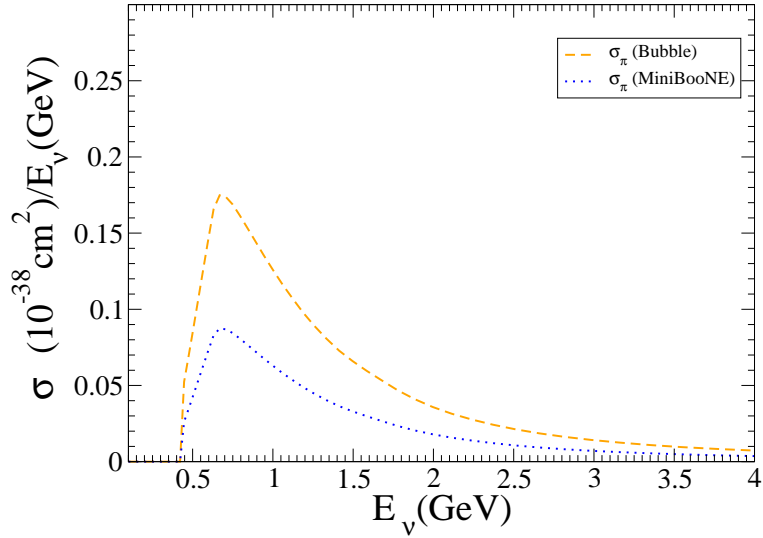}
\caption{ Results for the fits  to resonant $\pi^{0}$ production due to the reaction $\nu_{\mu}+N\rightarrow \mu+N'+\pi^{0}$, when  apply the kinematic cuts from Eq.~(\ref{pwindow}).}
\label{fig:vincpi0}
\end{center}
\end{figure}

It is clear  that an unified picture for resonant (neutral) pion production due to neutrino interactions in a nuclear medium does not appear and should be theme of intense discussion for both theoretical and experimental point of view and FSI are important component of such  subject. Hence, we limit ourself to use the cross-sections shown in lower panel of Fig.~\ref{fig:oldnew} for old and new data. In Fig.~\ref{fig:vincpi0} we show  the results for $\sigma/E_{\nu}$  resonant neutral pion production cross section  over neutrino energy as function of $E_{\nu}$  when   the cuts from  Eq.~(\ref{pwindow}) are applied.  We see that cuts severely reduces the cross-section at high energies and only the few GeV region is in practice relevant.
As a final comment about cross-sections we stress here that because of limits from  Eq.~(\ref{pwindow}) translates to the double differential cross-section, not only the integral of it, is necessary to perform the  proton decay background calculation.

%=================================================================== 
\section{Standard neutrino oscillations in atmosphere} 
\label{sec:SO}

\begin{figure}[h]
\hspace{-1.cm}\includegraphics[scale=0.33,angle=0]{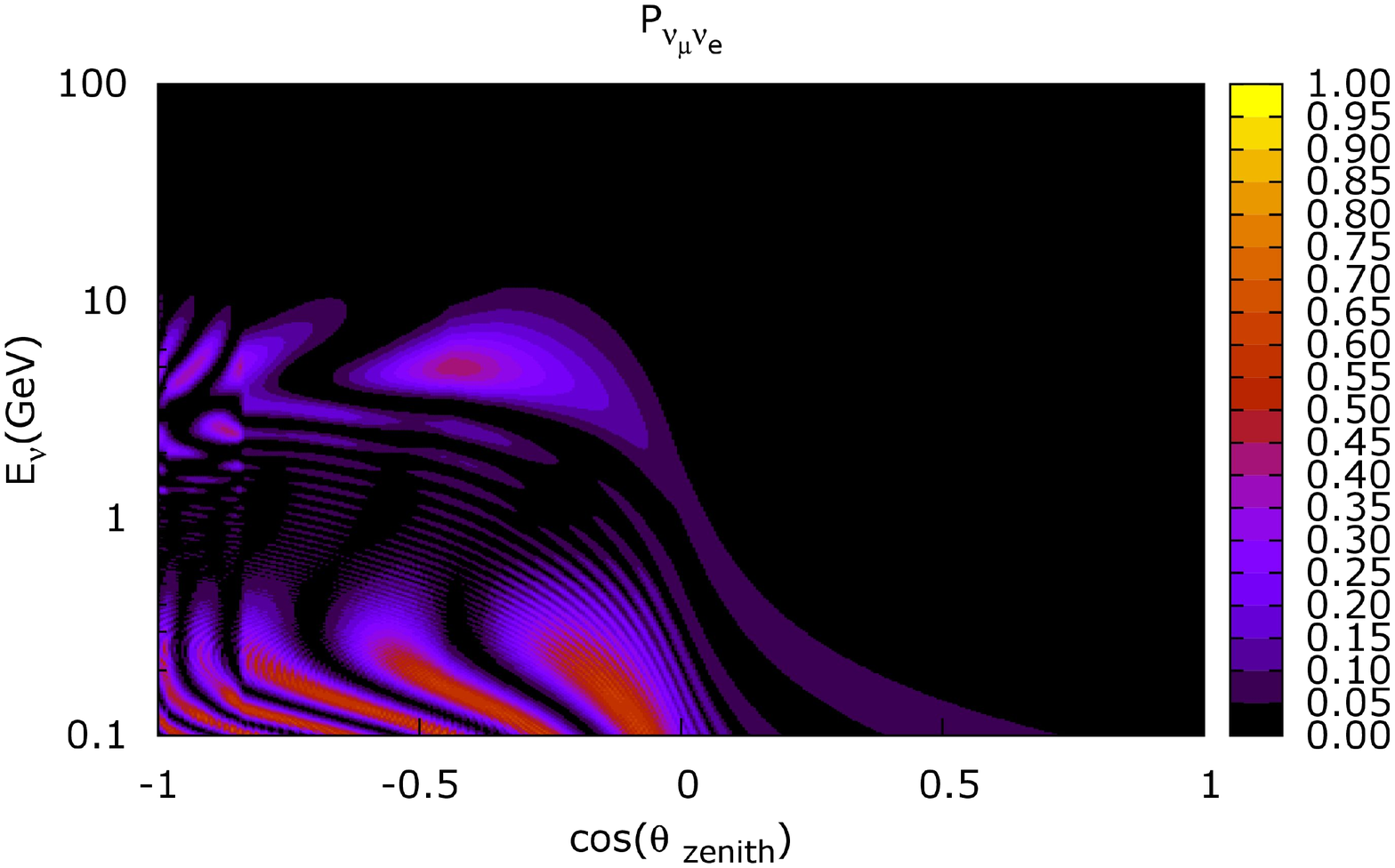}
\hspace{-1.4cm}\includegraphics[scale=0.33,angle=0]{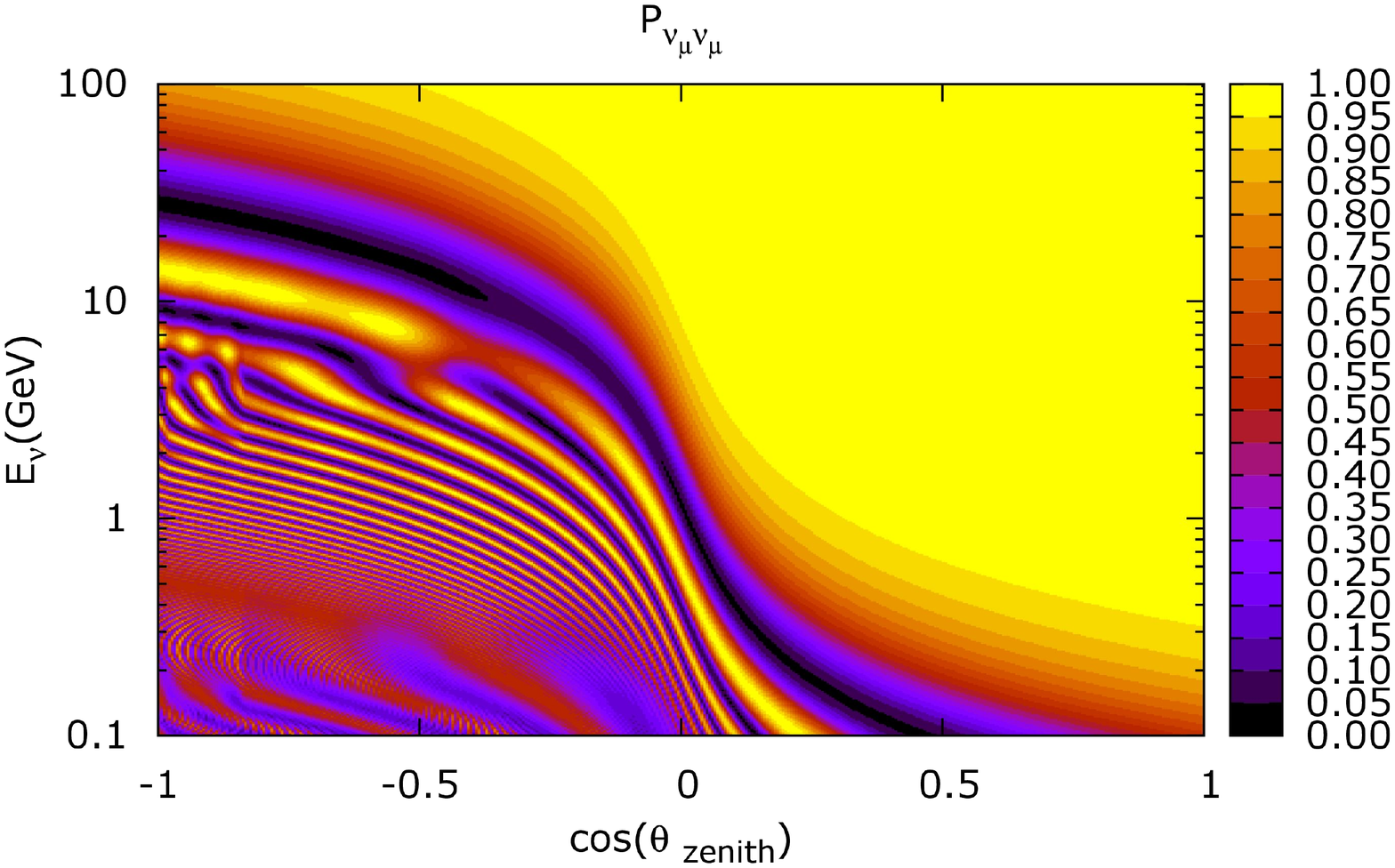}
\vspace{-2.cm}

\hspace{-1.cm}\includegraphics[scale=0.33,angle=0]{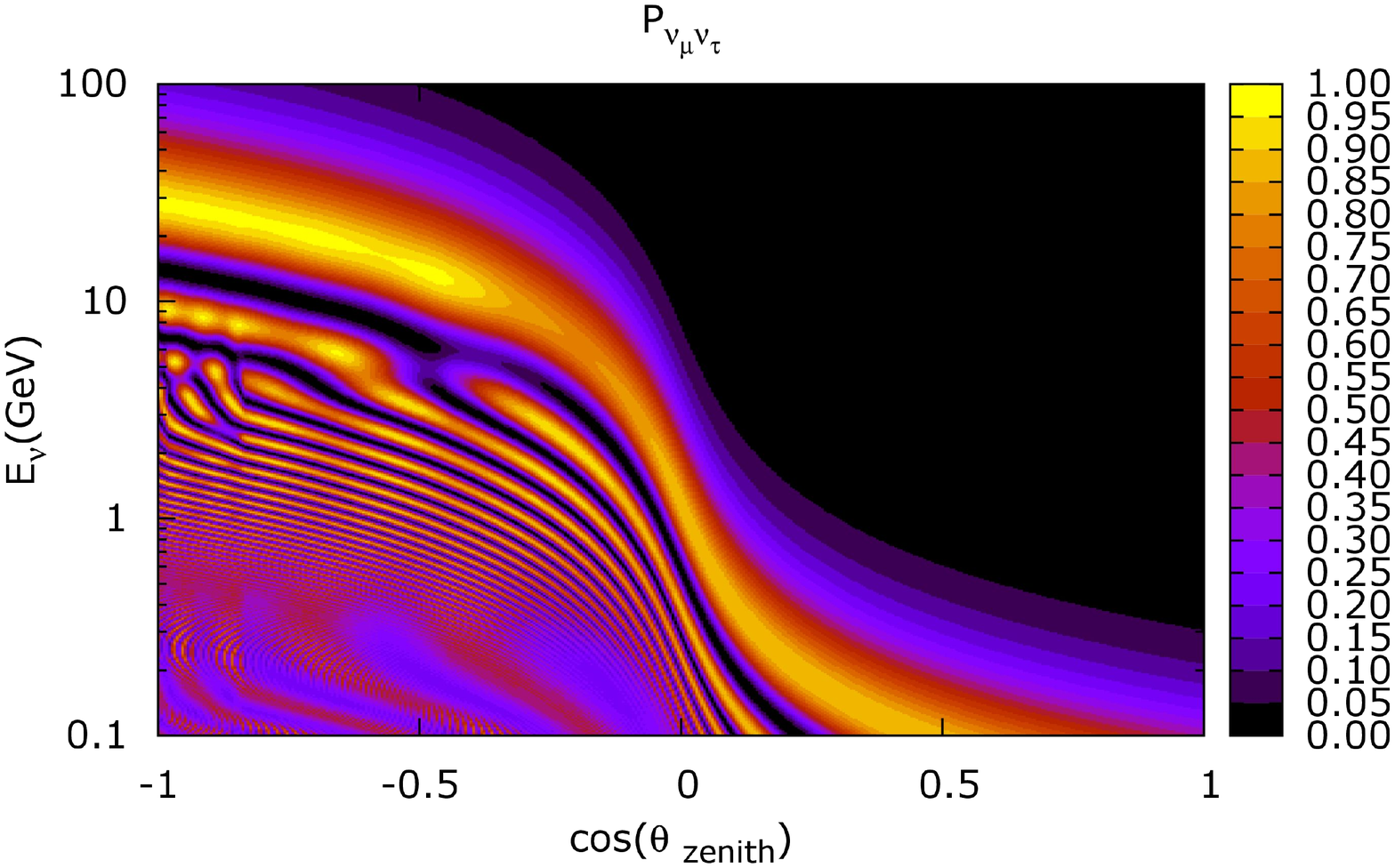}
\hspace{-1.4cm}\includegraphics[scale=.33,angle=0]{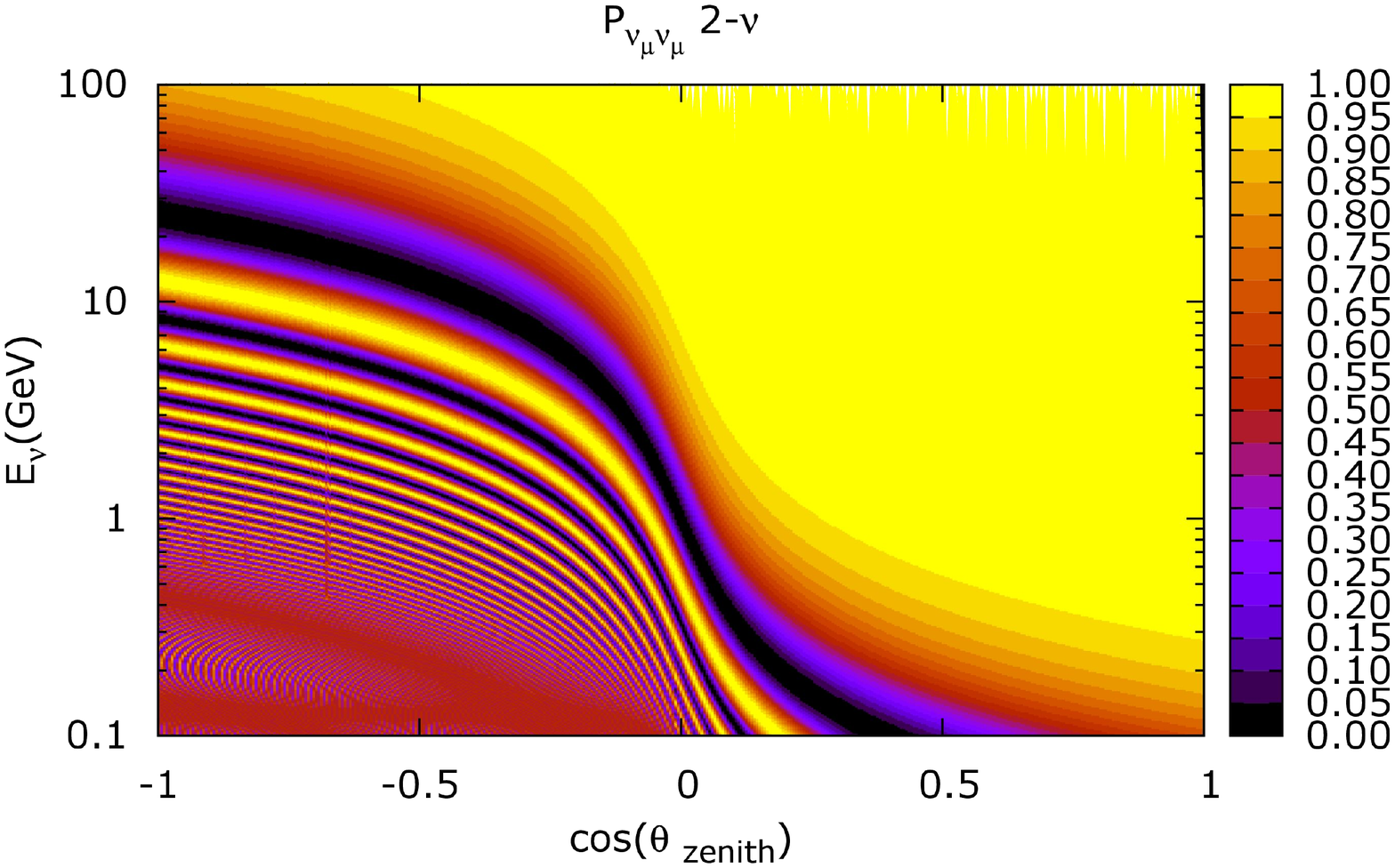}
\caption{Oscillograms for $P_{\nu_{\mu}\nu_{\alpha}}$. The upper-left, upper-right and  down-left  panels are respectively 
$P_{\nu_{\mu}\nu_{e}}$, $P_{\nu_{\mu}\nu_{\mu}}$, $P_{\nu_{\mu}\nu_{\tau}}$ solutions from Eq.~(\ref{SCH}). The  down-right panel is $P_{\nu_{\mu}\nu_{\mu}}$
as given from Eq.~(\ref{pmumu}). }
\label{SO}
\end{figure}

To include the full three-neutrino oscillation probability in atmospheric neutrinos we solve the Schrodinger-like equation  for the complete Hamiltonian, including the standard matter potential due to CC interactions between neutrinos and electrons inside the Earth, described by the potential $V(r)$~\cite{Mikheyev:1989dy,Wolfenstein:1977ue}. Hence, the complete three-flavor neutrino time-evolution equation is 
\begin{eqnarray}
i\dfrac{d \nu_\alpha }{dt} & = & \left[ \dfrac{1}{2p_\nu}U M^2 U^\dagger +V(r)\right]_{\alpha\beta}\nu_\beta,
\label{SCH}
\end{eqnarray}
\noindent where $\alpha=e,\mu,\tau$,  $U$ is the PMNS matrix~\cite{ref:PDG}, and $M^{2}$ is written in terms of the squared of mass eigenstates, $\Delta m^{2}_{ij}\equiv m_i^2-m_j^2$. The $M$ and $V$ matrices in Eq.~(\ref{SCH})  can written as
\begin{eqnarray}
 M^2  & = & {\rm diag}(0,\Delta m_{21}^2,\Delta m_{31}^2)\ , \\
V(r) & = & 2\sqrt{2}G_F\ N_e(r)\ {\rm diag}(1,0,0)\ ,
\label{mm}
\end{eqnarray}
\noindent where $G_F$ is the Fermi constant and $N_e(r)$ is the electron density profile of the Earth  from PREM model~\cite{Dziewonski:1981xy}. In this work we adopt the values of $\Delta m^{2}_{21}=7.4\times 10^{-5}$~eV$^{2}$, $\Delta m^{2}_{31}=2.47\times 10^{-3}$~eV$^{2}$, $\theta_{12}=33.36^{0}$, 
$\theta_{13}=8.66^{0}$, $\theta_{23}=40.00^{0}$~\cite{ref:PDG}.  Also, an useful approximation is the two neutrino oscillation limit  in vacuum, given by 
\begin{equation}
P_{\nu_{\mu}\rightarrow {\nu}_{\mu}}=1-\sin^{2}(2\theta_{23})\sin^{2}\left(1.27 \Delta m^{2}_{32}\dfrac{L}{E_{\nu}}\right)~,
\label{pmumu}
\end{equation}
where $L$ is given in km and $E_{\nu}$ in GeV. The rich pattern for muon neutrino survival and oscillation probabilities that emerges as solution of 
Eq.~(\ref{SCH}),  namely $P_{\nu_{\mu}\rightarrow \nu_{e}}$, $P_{\nu_{\mu}\rightarrow \nu_{\mu}}$, $P_{\nu_{\mu}\rightarrow \nu_{\tau}}$ are shown  through the oscillograms of  Fig.~\ref{SO}. The vertical axis is the neutrino energy $E_{\nu}$ and the horizontal one is the zenith angle of atmospheric neutrinos strike the detector, as counted from the nadir at detector position. 
As pointed previously, the atmospheric neutrino problem was identified as an reduction of $50\%$ in upward atmospheric neutrinos counted in SK when compared with theoretical predictions while for the downward direction the prediction and experimental results do math~\cite{ref:SK-evidence}. Such effects can be now perfectly understood in presence of oscillatory flavor transition as solution of Eq.~(\ref{SCH}). Given the average atmospheric neutrino energy around 1.0~GeV, the reduction in upward muon events is due to the higher distance traveled by neutrinos, while downward neutrinos which travel  shorter distance, have no enough space to oscillate. At such energies, for $\cos(\theta_{\rm zenith})=-1$ the oscillatory pattern in $P_{\nu_{\mu}\rightarrow \nu_{\mu}}$ is averaged out.
Hence  a deficit of $\approx 50\%$ also implies that the mixing angle be close to $45^{0}$. A noticeable fact is that the value of $\Delta m_{32}^{2}$ is in atmospheric neutrino experiments constrained by the region of intermediate values of $\cos(\theta_{\rm zenith})$,  but not at $\cos(\theta_{\rm zenith}) \rightarrow -1$, since at this region  oscillations are average out. As can be seen in oscillograms, $P_{\nu_{\mu} \rightarrow \nu_{e}}$ has significant magnitude ($\approx 0.6$) only at very narrow energy region and $P_{\nu_{\mu} \rightarrow \nu_{\mu}}$ and $P_{\nu_{\mu}\rightarrow \nu_{\mu}}$ mirror each other.  For comparison in the down-left panel we show  the results from 
Eq.~(\ref{pmumu}) for $P_{\nu_{\mu}\rightarrow \nu_{\mu}}$ for the same parameters. In the energy region we are interested in the  effects due to matter potential are not dominant. This fact combined with the known smallness of $\theta_{13}$~\cite{ref:dayabay} results that Eq.~(\ref{pmumu}) should be used in a first approximation to neutrino oscillations. Nevertheless, we use the complete numerical solution in all our calculations hereafter.  
Also,  as for energies around $E_{\nu}=1.0$~GeV, the ratio between muon and electron neutrinos produced in atmosphere is of order of two~\cite{ref:Honda:2006qj}, the effects of   $P_{\nu_{\mu} \rightarrow \nu_{e}}$ should play some role in determination of atmospheric neutrino background to proton decay in channel $p\rightarrow e+\pi^{0}$. 

%======================================================================
\section{Atmospheric neutrino background }
\label{sec:Natm}

 In this section we specify our procedure to calculate the number of {\it muon-like} events due to atmospheric neutrino flux that can be misidentified  with proton decay signal in the channel defined in  Eq.~(\ref{eq:pdec01}).  The charged current (CC)  neutrino-nucleon interaction is the dominant processes at atmospheric scale ($E_{\nu}\approx 0.1-100.0$~GeV) in such way that we simply disregard  all other processes, such as neutral current (NC) neutrino-nucleon, and neutrino-electron processes. Moreover, only the the resonant neutral pion production due to neutrino-nucleon charged-current scattering is responsible for  the background production in the case we are interested. The desired relation is given in  Eq.(\ref{pires1}), and the details about it are given in Sec.(\ref{sec:cs}). Generally, the number of atmospheric neutrino events is the integral of the product of  differential cross-section  with atmospheric neutrino flux and oscillation probability.  We define the number of muon-like events in the detector as
\begin{eqnarray}
N&=&n_{\rm tar}\epsilon T\int^{1}_{x_{0}}dx\int^{1}_{-1}d\cos(\theta_{\nu})\int^{2\pi}_{0}d\phi_{\nu}\int^{E_{\nu,f}}_{E_{\nu,0}}dE_{\nu}\int^{E_{l,f}}_{E_{l,0}}
dE_{l}\nonumber \\
&\times& \left[\sum_{\bar \nu, \nu}\phi_{\nu_{\mu}}(E_{\nu},\theta_{\nu},\phi_{\nu}) P_{\nu_{\mu}\rightarrow\nu_{\mu}}\frac{d\sigma_{\nu_{\mu}}}{dE_{\nu}dE_{l}dx}\right.\nonumber\\
&+&\left. \sum_{\bar \nu, \nu}\phi_{\nu_{e}}(E_{\nu},\theta_{\nu},\phi_{\nu}) P_{\nu_{e}\rightarrow\nu_{\mu}}\frac{d\sigma_{\nu_{\mu}}}{dE_{\nu}dE_{l}dx}\right]~,
\label{neve}
\end{eqnarray}
where $\epsilon=45\%$ is the detector efficiency for such process~\cite{ref:rubia},   $t=3.15\times 10^{8}$~s is the number of seconds in  $10$ years period of data taken.  Initially we assume that the DUNE far detector is a $40.0$ ~kton of pure liquid argon(Z$=18$, A$=40$). This implies $\approx 1.2\times 10^{34}$  protons in the effective volume. Also, in Eq.~(\ref{neve}), $\phi_{\nu}$ is the atmospheric neutrino or anti-neutrino flux as obtained from Honda~\cite{ref:Honda:2006qj}. The uncertainties in such fluxes in the overall normalization and in the angular shape are respectively of order of  25\%  
and 5\%~\cite{ref:rubia}.  These fluxes must be  multiplied for the standard neutrino flavor oscillation probability as defined in Sec. (\ref{sec:SO}), $P_{\nu_{\alpha}\rightarrow \nu_{\beta}}$, to count the right flavor of neutrinos that interacts with detector.
 
Now we apply the cuts necessary to select from atmospheric muon neutrino interactions the events that counts as background. Firstly note that in the 
Super-Kamiokande neutrino detector, the proton at final state is usually below the threshold for Cherenkov radiation emission, and so, is not detected. In liquid argon, the threshold for protons to be detected is a track about $80$~cm  \cite{ref:lqdar40}. Hence, in this calculation we follow  Ref.~\cite{ref:rubia} and proceed the sum over neutrino and anti-neutrino signals.  

The differential resonant neutral pion production cross-section,  $\dfrac{d\sigma_{\nu}}{dE_{\nu}dE_{l}dx}$ is  defined in Sec.(\ref{sec:cs}). As pointed, there is a discussion about the nature and  importance of nuclear effects both in first vertex of neutrino interaction as well as for the (multiple) pion scattering inside nuclear matter that are until today model dependent. Hence we limit ourself to apply the formalism from \cite{Fogli:1979cz} to describe  older~\cite{ref:barish} and newer~\cite{AguilarArevalo:2010zc} data.  We understand that results from  such calculation configure  respectively upper and lower limits on this observable and hope that an unified picture of double-differential resonant neutrino cross-section appears in near future. In this sense, the values  we found in Sec.~(\ref{sec:results}) should be understood as an error band, or a  guide-line for the  true value DUNE will find. 

Once cross-section is addressed, we limit the possible values of $E_{\mu}$ using Eq.~(\ref{pwindow}). Also, as pointed in Sec. \ref{sec:bkgd}, only $\approx 38\%$ of produced neutral pion follow in the energy window defined in  Eq.~(\ref{pwindow}). The last cut we include is the direction between muon and pion be back-to-back. In this calculation we recognize that whatever is the muon direction emission, the detector will have a finite resolution on this direction $\delta A_{\mu}$. For protons at rest, the event induced by atmospheric neutrinos only will counted as background if  pion from $\Delta$ decay is emitted  in an angular window in  opposite direction  $\delta A_{\pi^{0}}$ . For protons with some small momentum, the relative direction between their decay products is no longer back-to-back and depends on proton momentum. In this case our approach is to assume whatever is the angle between the charged lepton and the neutral pion, the angular window to detect this particles do not change in magnitude. Hence, in both cases, the fraction of atmospheric neutrino background to proton decay  is given then by 
Eq~(\ref{neve}) multiplied by  the fraction of angular area of uncertainty in identification of emission directions relative to  the total one. We write this area as function of polar coordinates in detector frame ($\theta, \phi$), 
and assume that the angular uncertainty is of same magnitude $\theta_{\rm max}= \phi_{\rm max}$, 
leading to 
\begin{equation}
 \delta A = \frac{\theta_{\rm max}(1+\cos(\theta_{\rm max}))}{4\pi}
\end{equation}
Here we stress that $\delta A_{\mu}$ should be different from $\delta A_{\pi^{0}}$, and hence $\theta_{\rm max}$ is the angle related to the total area from the sum of $\delta A_{\mu}$ and  $\delta A_{\pi^{0}}$. An intrinsic  source for the angular resolution error is the Compton effect, which can not be avoided at fundamental level~\cite{ref:largo}.

\begin{figure}[h]
\begin{center}
\includegraphics[scale=0.4]{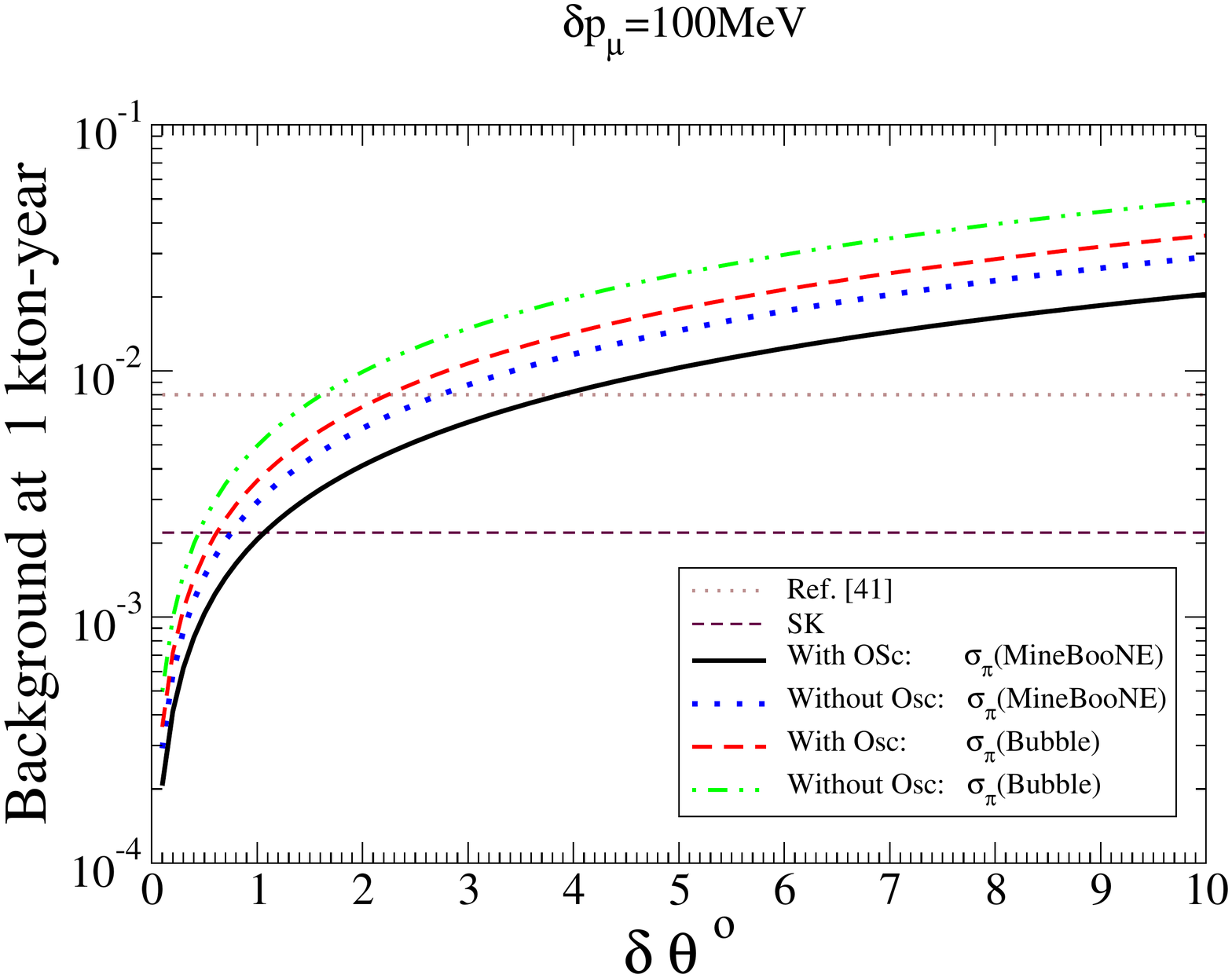}

\includegraphics[scale=0.4]{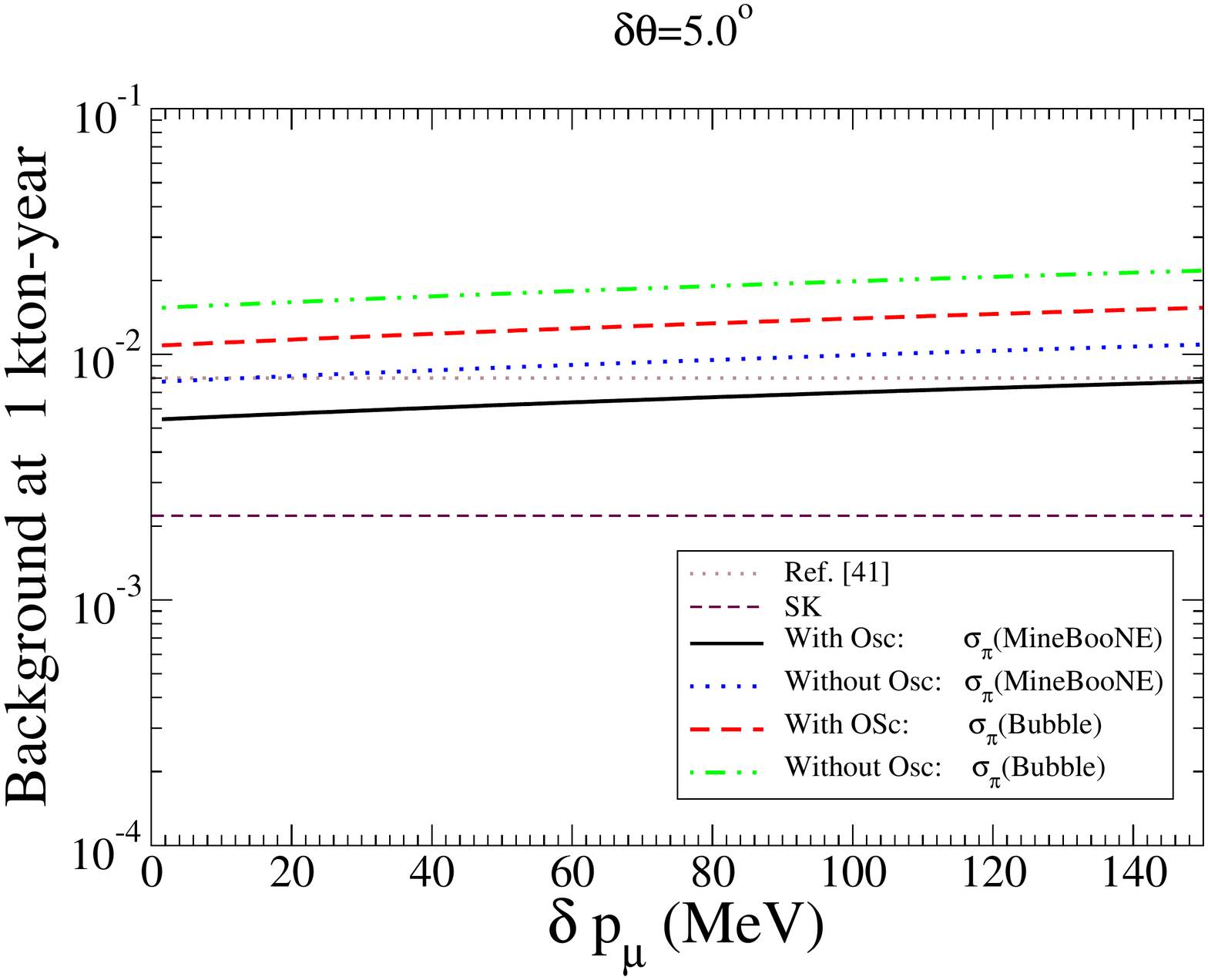}
\caption{The atmospheric neutrino background (number of events) for proton decay in DUNE for the  process given by Eq. (\ref{neve}). 
Upper Panel: As function of angular uncertainty, for $\delta {p}_{\mu}=100$~MeV. Lower Panel: As function of muon momentum uncertainty for $\delta \theta=5^{o}$. An averaged efficiency of $\epsilon=45\%$ was  applied in all cases. The prediction from SK~\cite{SK-pdecay} and DUNE~\cite{ref:rubia} are also shown.}
 \label{bkgd}
 \end{center}
 \end{figure}

We apply the above cuts on Eq.(\ref{neve}) and the results atmospheric neutrino background  as function of $\delta A$ are shown  in upper panel of Fig.~\ref{bkgd}.  We compare it with SK~\cite{SK-pdecay} results and  also  predictions for DUNE from~\cite{ref:rubia}.  Our calculations are for $1.0$ kton-year and we renormalize the references to allow direct comparison. We find that the angular resolution $\delta A$  has an important impact over the absolute background value. In a scenario with standard oscillation and using  $\sigma_{\pi}(MiniBooNE)$, the reduction in background is of a factor 2 when $\theta_{max} $ goes from $10^{0}$ to $5^{0}$ and of a factor 4.5 when it goes from $5^{0}$ to $1^{0}$. In that case however, the impact of standard oscillations is of order of $35\%$ at $\theta_{max}=1^{0}$ and of $40\%$ at $\theta_{max}=5^{0}$, and holds the same value for $\theta_{max}=10^{0}$. When we use $\sigma_{\pi}(Bubble)$ cross-section, the effect of include standard oscillation is of order of $46\%,40\%, 36\%$ at $\theta_{max}=1^{0},5^{0},10^{0}$.  In Fig.~\ref{bkgd}, lower panel,  the results for the background are shown as function of uncertainty in muon momentum. In presence of standard neutrino oscillations, as $p_{\mu}$ increases, the reduction on background due to use of MiniBooNe data fit is a factor that varies from $2.00$ to $2.14$ when compared with the older bubble fit.  On the other hand, assuming $\sigma_{\pi}(MiniBooNE)$ cross-section the reduction in background  due to inclusion of standard oscillation is of order of $40\%(30\%)$ for $\delta p_{\mu}=0(150)$~MeV. In next section we investigate how such reductions should improve the limits on proton lifetime. 

\section{Atmospheric neutrino background and limits on proton lifetime  }
\label{sec:results}

In this section we discuss how the changes in the background due to atmospheric neutrinos impacts the limits on the proton lifetime. We use Poisson distribution 
\begin{equation}
P(n, b+S)=\dfrac{e^{-\lambda}(\lambda)^{n}}{n!}.
\end{equation}
where  $b$ is the background, $S$ and $\lambda=b+S$ and $n$ is the expected signal. The probability of measure $n$, within a expectation value $\lambda=b+S$. For a given value of  $b$, the signal  $S$ we want for a significance of $ 90 \% $~C.L. , that means $\alpha =0.1$, is given by~\cite{ref:PDG,ref:rubia}
\begin{equation}
\dfrac{\sum^{n_{0}}_{n=0}P(n,b+S)}{\sum^{n_{0}}_{n=0}P(n,b)}=\alpha=0.1, 
\label{eq:poisson}
\end{equation}
where $n_{0}$ is the closest integer number to $b$. This means we have $90\%$ of C. L. that we see the signal and not a fluctuation from background. In practice, Eq.~(\ref{eq:poisson}) turns in  $S=S(b)$,  or equivalently, in a functional of the net detector  mass and time of exposition,  $S=S({\rm kton-yr}~\epsilon)$, where kton-yr is the exposition in kton-years and  $\epsilon$ is the overall efficiency.  It is important to say that for $400$ kton-year we found the expected $b$ is of order of $\approx 3.0-7.9$ events(See Tables~\ref{Tab:01} and  \ref{Tab:01A}). Hence, for low values of exposition, the fluctuations in such very low background  are not negligible. In terms of formalism, the fact that $n_{0}$ is discrete (the closest integer to $b$) has some impact. In the exponential in Eq. (\ref{eq:poisson}), for  higher value of $(b+S)$, the smaller the ratio. Hence, if $b$ increases but not enough to change $n_{0}$,  this increment in background implies in a smaller signal to reach the same confidence level. However, when the increases in $b$ is big enough to increase $n$ by one unit, then the sum in 
Eq.~ (\ref{eq:poisson}) has one more term. 

Once $S$ is known, the proton lifetime should be written as~\cite{ref:rubia}:
\begin{equation}
\frac{\tau}{\beta}(\rm{ years})>\dfrac{2.7}{S({\rm kton}\epsilon)} \rm {kT} \epsilon ~10^{32},   
\label{eq:plf}
\end{equation}
where $2.7 \times 10^{32}$ is the number of protons in one kton of  $^{40}$Ar  and $\beta$ is the branching ratio to the channel. 

\begin{table}[htb] % [htb]-> here, top, bottom
\small
\centering   
\caption{Results for the atmospheric neutrino background from Eq.~(\ref{neve}) with (SO) and without (NSO) the inclusion of standard neutrino oscillations. 
Data are normalized to 40 kton of fiducial volume and 10 years of data-taken. Here we use  $\sigma_{\pi}(Bubble)$. We set $\delta \theta=5^{o}$.}
\begin{tabular}{|c|c|c|c|c|c|c|c|c|} % c=center, l=left, r=right
\hline
$\delta p_{\mu}=100$~MeV & With Osc.  & Without Osc.   &  $\delta p_{\mu}=70$~MeV &   With Osc.  & Without Osc.   \\
\hline
Background& 5.62& 7.92& Background& 5.21 &7.42\\
\hline
S&  5.71  & 6.13&S & 5.06& 5.59\\
\hline
S+B&  11.33& 14.95 &S+B& 10.27& 13.01 \\
\hline
{\rm $\tau $(years)}&$8.50\times 10^{33}$ & $7.91\times 10^{33}$ & T (years) & $9.58\times 10^{33}$ & $8.68~10^{33}$ \\
\hline
\end{tabular}
\label{Tab:01}
\end{table}

\begin{table}[htb] % [htb]-> here, top, bottom
\small
\centering   
\caption{Results for the atmospheric neutrino background from Eq.(\ref{neve}) with (SO) and without (NSO) the inclusion of Standard Neutrino 
Oscillations . Data are normalized to  40 kton and 10 years of data-taken. Here we use  $\sigma_{\pi}(MiniBoone)$. We set  $\delta \theta=5^{o}$.}
\begin{tabular}{|c|c|c|c|c|c|c|c|c|} % c=center, l=left, r=right
\hline
 $\delta p_{\mu}$=100~{\rm MeV}&With Osc.&Without Osc.&$\delta p_{\mu}=70$~MeV& With Osc.&Without Osc.\\
\hline
Background& 3.30& 4.64  & Background&3. & 4.36\\
\hline
S&    4.31 & 5.36 &S  & 4.4& 4.64 \\
\hline
S+B& 7.61&  10.00  &S+B& 7.4 & 9.00 \\
\hline
{\rm $\tau$(years)}& $1.13\times 10^{34}$& $9.06\times 10^{33}$ & T(years) & $1.09\times 10^{34}$ & $1.04\times 10^{34}$ \\
\hline
\end{tabular}
\label{Tab:01A}
\end{table}

In Table (\ref{Tab:01}) (Table~(\ref{Tab:01A})) are shown our results for proton lifetime when  we use   $\sigma_{\pi}(Bubble)$ ($\sigma_{\pi}(MiniBooNE)$) data. Both are calculated for $400$ kton-years. From Tabs. \ref{Tab:01} and \ref{Tab:01A} we observe that the reduction on atmospheric muon-like neutrino background when we include standard neutrino oscillations is of order of   $40\%$.  Also  $\sigma_{\pi}(MiniBooNE)$  implies in a reduction of at least $70\%$ in background when compared with the case where $\sigma_{\pi}(Bubble)$ is used. Combined, such  improvements leads to an reduction of a factor $2.40~(2.47)$ in the background for $\delta p_{\mu}=100(70)$ MeV. However, as pointed in the discussion of Eq.(\ref{eq:poisson}), an reduction in the background does not imply in a strictly proportional  improvement of proton lifetime limit, since it  depends on $S$, not on $b$ directly. Even in this case,  the inclusion of both Standard Oscillations and cross-sections tunned with MiniBooNE  data leads to an improvement of $42\% ~ (44\%)$ in proton lifetime at $90\%$ C. L. for  $\delta p_{\mu}=100 ~ (70)$ MeV  and at exposition of $400$ kton-years. We verified that the modifications due only to Standard Oscillations  are of order of $7\%~(10\%)$ when we use bubble cross-section and $\delta p_{\mu}=100~ (70)$~MeV  . When  $\sigma_{\pi}(MiniBooNE)$  is used, this reduction is of $ 24\%~(5\%)$ for the same values of $\delta p_{\mu}$. The reduction of impact due to standard neutrino oscillation in latter case is expected since, due to lower phase-space and cross-section combined, the value of $b$ is lower than the former case.

\begin{figure}[hbt]
\begin{center}
\includegraphics[scale=0.4]{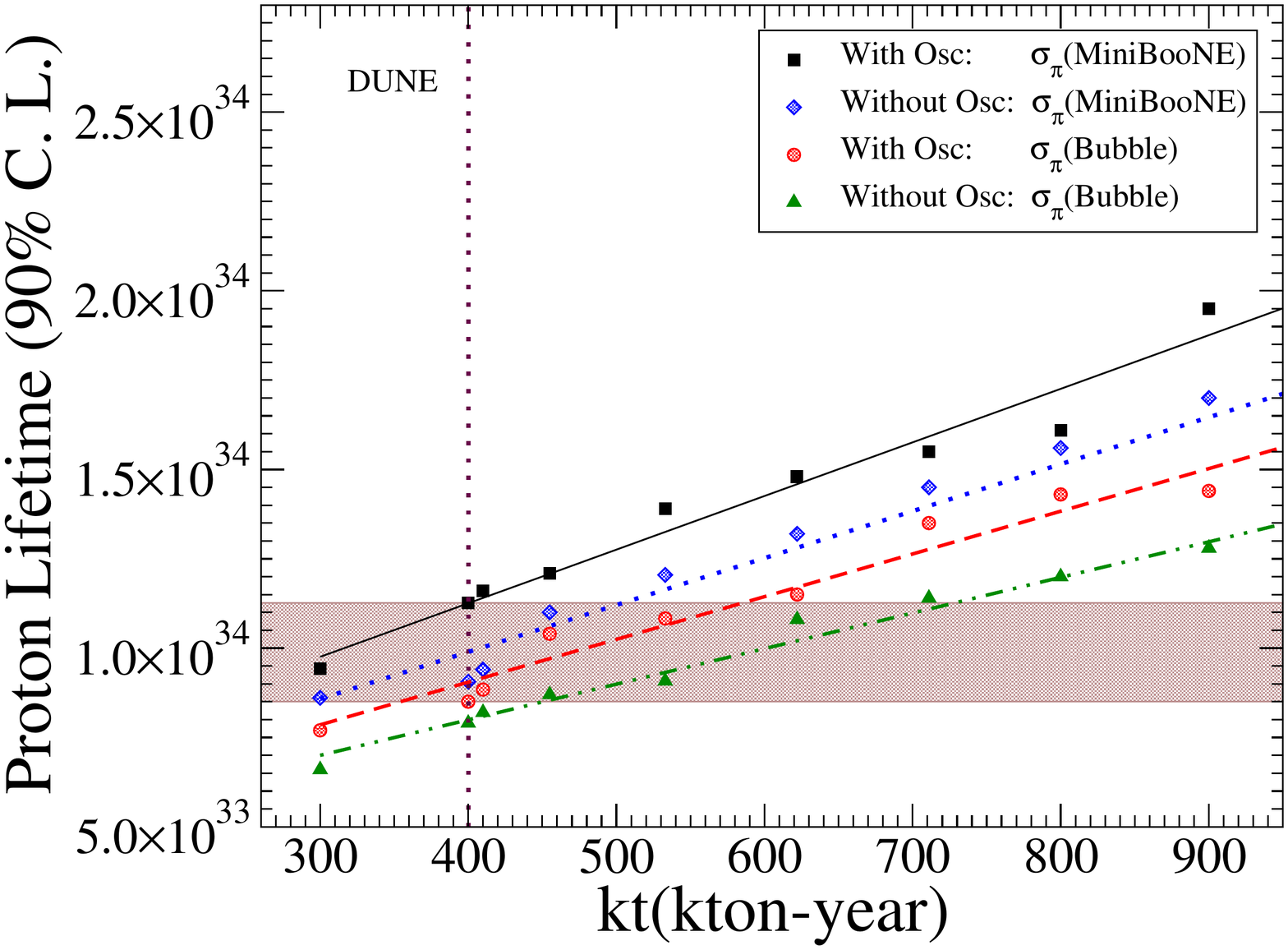}\vspace{-0.5cm}

\includegraphics[scale=0.4]{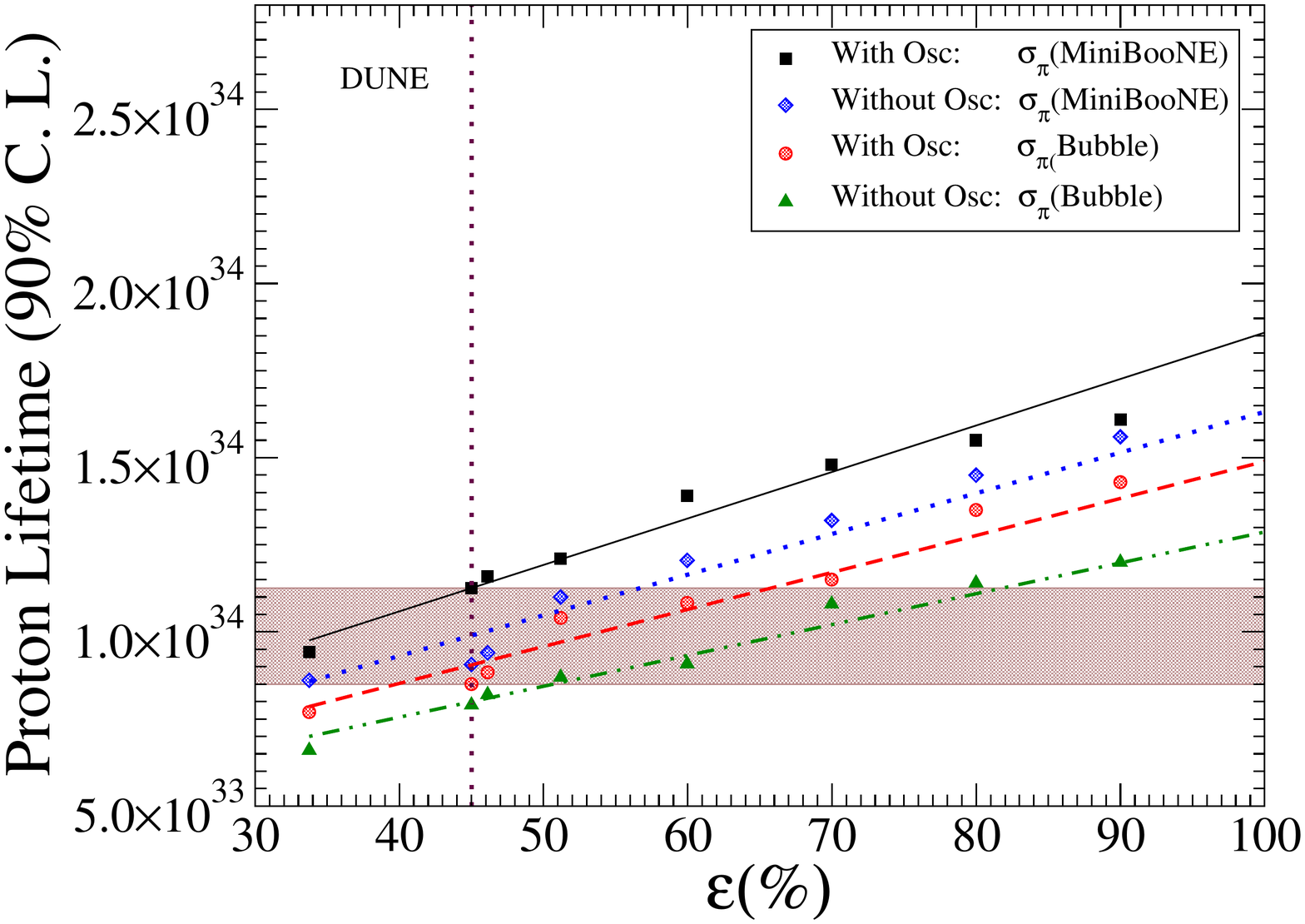}
\caption{Upper Panel: Impact of inclusion of standard neutrino oscillations in proton lifetime limit as function of exposition in kton-year for the resonant cross-section as indicated in plot. We set  $\delta p_{\mu}=100$~MeV and $10$ years of data taken. Lower Panel: The same as function of overall efficiency.  In all plots, the fluctuations in points are due to Poisson distribution of background, as given in 
Eq.~(\ref{eq:poisson}). We also present the correspondent tendency line to each case. The shaded area is the error band due to use upper or lower bound on cross-section, both with the inclusion of standard oscillations. 
}
 \label{fig:limit_kt}
 \end{center}
 \end{figure}

We also verify how such limits change as a function of exposition kton for fixed efficiency $\epsilon=45\%$. In Fig.~\ref{fig:limit_kt}, upper panel, we show our predictions for proton lifetime as function of exposition, for the cases with and without Standard Oscillations and for both MiniBooNE and Bubble cross-sections. However we understand that an exposition of order of a thousand of kton-years is not a realistic possibility. Hence, in lower panel of Fig.~\ref{fig:limit_kt} we show the region which could be reached with the same 40 kton detector and 10 years of exposition, in case of future improvements in the overall efficiency. Apart from fluctuations, $b$ and $\tau$ should scale with  ${\rm kton}\epsilon$, and an increase of $10\%$ in kton or $\epsilon$ has in practice the same effect.  
 
\section{Conclusions}
\label{sec:conc}

We study the impact of four systematics in the background to proton decay in the channel $p\rightarrow \mu+\pi^{0}$, namely,  the overall efficiency, the muon reconstruction energy resolution, the resonant neutral pion cross-section and the neutral pion angular resolution. We also include standard neutrino oscillations to calculate the reduction of around $40\%$ of the background to proton decay at a DUNE-like detector due to atmospheric neutrinos.  The  pion resonant cross-section tuned with MiniBooNE data also reduces the background when compared with old bubble data at least $70\%$. However, as pointed in Sec~\ref{sec:results}, as  the absolute value of background is too small, a reduction of it does not implies a strictly  proportional increment of proton lifetime expectation.  We found that $\sigma_{\pi}(MiniBooNE)$ cross-section improves proton lifetime limit  in $40\%$ and the inclusion of standard oscillations improves such limit in $\approx 24\%$ for the present limit of proton decay~\cite{ref:PDG}.  Combined, such features reduced background in a factor of $2.4$.  For a realistic detector of $40$~kton taking data for 10 years, we obtained a  range in limit of  proton lifetime of $7.9 \times 10^{33}\le \tau\le 1.1 \times 10^{34}$~s. This represents an improvement factor of $3.7-5.4$ in comparison  to the values obtained in Ref.~\cite{ref:PDG}.

A final comment is in order. One has to seek for any possible improvement of the efficiency of any detector devoted to set a limit to proton lifetime. An example of a very promising device to achieve such aim is ARAPUCA, which a photon trap device consisting in a polymer box that should  be installed  around every charge CCD (charge coupling device)  cam in DUNE~\cite{ref:arapuca}. The basic idea is that once the photon enter in the device it becomes entangled by reflections in the device   internal walls.  Hence, after a number of reflection, such photon is eventually absorbed by the photomultiplier. This should improve the detector overall efficiency. An analytical estimative of the gain for each photomultiplier due to ARAPUCA device can be found in Ref.~\cite{ref:etore}. ARAPUCA should increases by a factor at least $10$ the photomultiplier (PMT)  efficiency. The overall detector efficiency will be dependent, between other factors,  of total number and geometrical disposition of PMTs. As pointed in the right panel of Fig.~\ref{fig:limit_kt},  we show the sensitivity of limit on proton lifetime on such improvement in overall efficiency. Here we also argue that another possible feature of ARAPUCA may be to  reduce systematics as  the  error on muon momentum $p_{\mu}$ reconstruction, which also will have positive impacts on the reduction background to proton decay.

%%%%%%%%%%%%%%%%%%%%%%
\begin{acknowledgments}
We thank to M. Diwan for the suggestion of this topic.  D.R.G. is thankful to  E. Segreto, L.F.G. Gonzales, R.D. de Souza and V.P.G. Gon\c{c}alves for the very enlightening  physical discussions. M.M.G. is thankful to FAPESP and CNPq for several financial supports. O.L.G.P. is  thankful for the support of FAPESP funding Grant  2014/19164-6, CNPq research fellowship 307269/2013-2 and  304715/2016-6 and for partial support of ICTP.

\end{acknowledgments}

 \appendix
 \section{Kinematics} 
 \label{apk}

In this work  we are interested in the background due to  atmospheric neutrinos to the proton decay signal for the decay reaction Eq~(\ref{eq:pdec01}). We write the total energy for the final decay state as
\begin{equation}
E^{2}_{\rm tot}=(m_{p}\pm \delta m)^{2}, 
\end{equation}
where $\delta m$ is the error band in the energy of final state due to the incapability of detector to reconstruct perfectly this quantity. Assuming that protons are at rest when decay, $\vec p_{p}=\vec 0$, from conservation of energy, $E_{p}=E_{\mu}+E_{\pi}$,  and applying 3-momentum conservation, $\vec p_{\mu}=-\vec p_{\pi}$, we have for $\vec p_{\mu}$
\begin{eqnarray}
|\vec p_{\mu}|&=&\frac{c}{2m_{p}}\sqrt{(m^{2}_{p}+m^{2}_{\pi}-m^{2}_{\mu})^{2}-4m^{2}_{\pi}m^{2}_{p}}~.
\label{pmu}
\end{eqnarray}
Now taking into account that protons are not free inside the detector, but bounded to other protons and neutrons in the nucleons. This means that the effective proton mass is not $m_{p}=937,49$~MeV/c$^{2}$, but must be calculated ~\cite{ref:testeSK}. When we assume $m_{p}=925$~MeV/c$^{2}$ in Eq.~(\ref{pmu}) we find $|\vec p_{\mu}|=446.4$~MeV/c. In this case
\begin{eqnarray}
E_{\mu}&=&\sqrt{m^{2}_{\mu}c^{2} +|\vec p_{\mu}|^{2}}=458.7~{\rm MeV}\quad E_{\pi}=\sqrt{m^{2}_{\pi}c^{2} +|\vec p_{\mu}|^{2}}=466.4~{\rm MeV}~.
\end{eqnarray}
In  general case, protons should have some initial momenta before decay. This initial momentum must be take into account in the kinematics of decay. An important consequence is that in the laboratory frame, the final momentum of muon from proton decay is now dependent of the relative direction between the initial proton momentum $ p_{p}$ and the muon momentum, $p_{ \mu}$. As before, this can be obtained from the calculation of the invariants $p^{a}_{\mu}p_{a,\mu}$ and $p^{a}_{p}p_{a,p}$ in the frames of proton rest and the laboratory. After some algebra the desired relation is~\cite{goldanski}.
\begin{equation}
p_{\mu}=\dfrac{m_{p}E^{*}_{\mu}p_{p}\cos\theta_{\mu}+E_{p}\sqrt{m^{2}_{p}p^{*^{2}}_{\mu} -m^{2}_{\mu}p^{2}_{p}\sin^{2}\theta_{\mu}}}{E^{2}_{p}-p^{2}_{p}\cos^{2}\theta_{\mu}} 
\label{pmuslab1}
\end{equation}
where $p^{*}_{\mu}$ and $E^{*}_{\mu}$ are the muon momentum and energy in the proton rest frame. Also $\theta_{\mu}$ is the emission angle of 
the muon with respect to the proton momentum in the laboratory frame.  Clearly, the 
maximum (minimum) of $\vec{p}_{\mu}$ occurs when the muon emission is in the same (opposite) direction of proton's initial momentum,  $\theta_{\mu}= 0(180^{0})$.  In this case Eq.~(\ref{pmuslab1}) reduces to
\begin{eqnarray}
p_{\mu}^{\rm max (min)}=\frac{\pm E^{*}_{\mu}p_{p}+E_{p}p^{*}_{\mu}}{m_{p}}=588.7(340.7)~{\rm MeV}
\label{pmuslab0}
\end{eqnarray}
Here we assume the maximum momentum for the proton to be $p_{p}=250$~MeV.

\end{document}